\documentclass[twocolumn,tighten]{aastex63}
\usepackage{graphicx}
\usepackage{gensymb}
\usepackage{amsmath}
\usepackage{amsmath}

\def\kms{$\rm km~s^{-1}$}

\begin{document}
\title{The Spatial and Emission Properties of the Large
[O III] Emission Nebula Near M31}

\author[0000-0003-3829-2056]{Robert A.\ Fesen}
\affiliation{6127 Wilder Lab, Department of Physics and Astronomy, Dartmouth College, Hanover, NH, 03755, USA}

\author[0000-0003-2379-0474]{Stefan Kimeswenger}
\affiliation{Universit{\"a}t Innsbruck, Institut f{\"u}r Astro- und Teilchenphysik,  Technikerstr. 25\/8, 6020 Innsbruck, Austria}
\affiliation{Universidad Cat\'olica del Norte, Instituto de Astronom{\'i}a,  Av. Angamos 0610, Antofagasta, Chile}    

\author[0000-0002-4594-9936]{J. Michael Shull} ,
\affiliation{Department of Astrophysical and Planetary Sciences and CASA, University of Colorado, 389-UCB, Boulder, CO 80309, USA}

\author[0000-0002-7855-3292]{Marcel Drechsler}
\affiliation{\'Equipe StDr, B{\"a}renstein, Feldstraße 17, 09471 B{\"a}renstein, Germany}

\author[0000-0002-3172-965X]{Xavier Strottner}
\affiliation{\'Equipe StDr, Montfraze, 01370 Saint Etienne Du Bois, France}

\author{Yann Sainty}
\affiliation{54000 Nancy, Lorraine, France}

\author{Bray Falls}
\affiliation{Sierra Remote Observatories, 42120 Bald Mountain Road, Auberry, CA, 93602, USA}

\author{Christophe Vergnes}
\affiliation{Various Amateur Observatory Sites, Lorraine, France}

\author{Nicolas Martino}
\affiliation{Various Amateur Observatory Sites, Lorraine, France}

\author{Sean Walker}
\affiliation{MDW Sky Survey, New Mexico Skies Observatory, Mayhill, NM, 88339, USA}

\author{Justin Rupert}
\affil{MDM Observatory, Kitt Peak National Observatory, 950 N. Cherry Ave., Tucson, AZ 85719, USA}

\begin{abstract}

Drechsler et al.\ (2023) reported the unexpected discovery of a $1.5\degr$ long [\ion{O}{3}] emission nebula $1.2\degr$ southeast of the M31 nucleus.  Here we present additional images of this large emission arc, called SDSO, along with radial velocity and flux measurements from low-dispersion spectra. Independent sets of [\ion{O}{3}] images show SDSO to be composed of broad streaks of diffuse emission aligned NE-SW. Deep H$\alpha$ images reveal no strong coincident emission suggesting a high [\ion{O}{3}]/H$\alpha$ ratio. We also find no other [\ion{O}{3}] emission nebulosity as bright as SDSO within several degrees of M31 and no filamentary H$\alpha$ emission connected to SDSO.  Optical spectra taken along the nebula's northern limb reveal [\ion{O}{3}] $\lambda\lambda$4959,5007
emissions matching the location and extent seen in our [\ion{O}{3}] images. The 
heliocentric velocity of this [\ion{O}{3}] nebulosity is $-9.8 \pm 6.8$ km s$^{-1}$ with a peak surface brightness of $(4\pm2) \times 10^{-18}$ erg s$^{-1}$ cm$^{-2}$  arcsec$^{-2}$ ($\sim$0.55 Rayleigh). 
We discuss SDSO as a possible unrecognized supernova remnant, a large and unusually nearby planetary nebula,  a stellar bow shock nebula, or an interaction of M31's outer halo with Local Group circumgalactic gas. We conclude that galactic origins for SDSO are unlikely and favor instead an extragalactic M31 halo--circumgalactic cloud interaction scenario, despite the nebula's low radial velocity.  We then describe new observations that may help resolve the nature and origin of this large nebulosity so close to M31 in the sky.

\end{abstract}
\bigskip

\section{Introduction}

Beginning with \citet{Minkowski1946,Minkowski1947,Minkowski1948}, 
\citet{Shajn1952}, \citet{Sharpless1953}, \cite{Gum1953,Gum1955},
and \citet{Morgan1955}, optical emission line surveys of the sky have been 
an especially effective means of identifying various types of emission nebulae such as  H~II regions, planetary nebulae (PNe), supernova remnants (SNRs), stellar wind-blown bubbles, and stellar outflows. The majority of these surveys have concentrated on detecting H$\alpha$ line emission along the Galactic plane. 
Recent surveys include the Virginia Tech Spectral-line Survey (VTSS; \citealt{Dennison1998}), the Southern H$\alpha$ Sky Survey Atlas (SHASSA; \citealt{Gaustad2001}),
the Wisconsin H$\alpha$ Mapper (WHAM; \citealt{Haffner2003}),
the AAO/UKST SuperCOSMOS H$\alpha$ survey \citep{Parker2005}, and the
Issac Newton Telescope Photometric H$\alpha$ Survey of the Northern Galactic Plane 
(IPHAS; \citealt{Drew2005})

The first comprehensive optical survey that included other nebular emission lines besides H$\alpha$ was the 1970's photographic survey of the Galactic plane by \citet{Parker1979}. This survey, using a small
commercial camera lens mounted ahead of a two-stage image intensifier,
generated moderately deep images in H$\beta$, 
[\ion{O}{3}] $\lambda$5007, H$\alpha$ + [\ion{N}{2}] $\lambda\lambda$6548,6583, and [\ion{S}{2}] $\lambda\lambda$6716,6731 line emissions.  These images quickly led to the discovery of several large PNe, Wolf-Rayet and OB star ring nebulae, and Galactic SNRs  
\citep{Gull1977,Kirshner1978,Gull1979,Blair1980,Fesen1981a,Fesen1981b,Fesen1983a,Fesen1983b,Chu1981,Bruh1981,Joy1982a,Joy1982b}. 

Like most emission line surveys, the \citet{Parker1979} survey was limited to the Galactic plane. However, with the advent of 
sensitive large-format CMOS  detectors, together with the development of 
high transmission (T$\sim95$\%) narrow passband 
filters (FWHM $\sim$ 30 \AA), small aperture telescopes are increasingly discovering new Galactic emission line nebulae, both large and small, and not just regions near the Galactic plane. The combination of wide fields of view and large pixel scales ($\geq2''$) offered by small telescopes has led to the discovery of dozens of new PNe \citep{Kronberger2014a,Kronberger2014b,LeDu2022}, low surface brightness stellar mass loss nebulosities \citep{Stefan2021}, investigations of faint stellar halos of spiral galaxies \citep{Abraham2014,Gilhuly2022},
an assortment of large-scale tidal structures around nearby massive galaxies \citep{Martinez2019},
and faint Galactic SNR filaments \citep{How2018,Fesen2021}.

In an unexpected discovery, \citet{Drechsler2023} detected an extremely faint $\simeq1.5\degr$ long [\ion{O}{3}] emission arc near M31 through the co-addition of hundreds of exposures obtained with small aperture telescopes (see Fig.~1). This emission arc, named the
Strottner-Drechsler-Sainty Object (SDSO), had not been reported previously
in any deep broadband or H$\alpha$ images of M31, and has no obvious emission counterpart in X-ray, UV, optical, infrared, or continuum radio surveys.
Here we describe more fully the instruments and data that led to the discovery of SDSO and the spatial and emission properties of this large [\ion{O}{3}] emission nebulosity near M31. We also discuss various Galactic and extragalactic interpretations of this emission arc.

\begin{figure*}[ht]
\begin{center}
\includegraphics[angle=0,width=18.0cm]{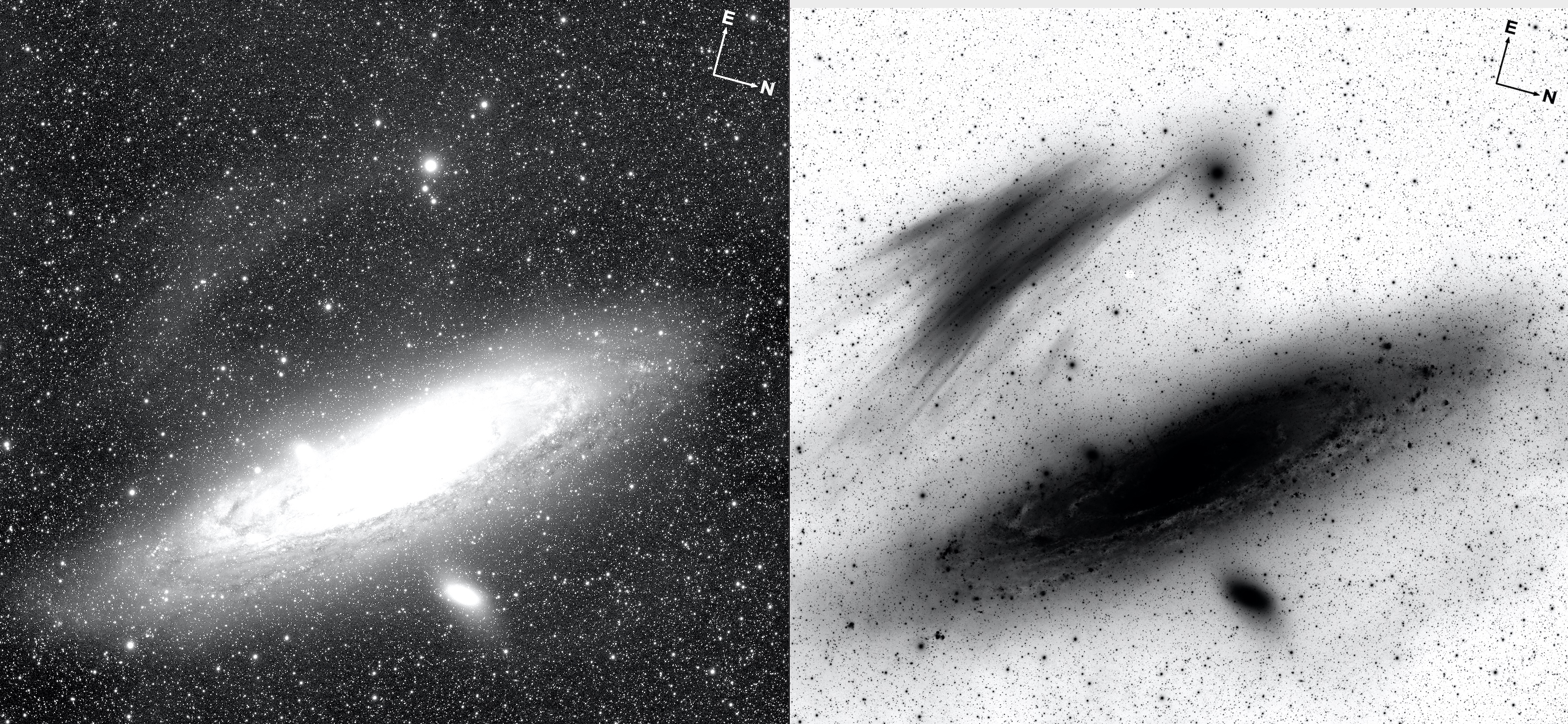} \\ 
\caption{ {\it{Left:}} Narrow passband 5007 \AA \ image, leading to the discovery of a large, very faint [\ion{O}{3}] emission nebula near M31. {\it{Right:}} Negative image after blue continuum subtraction and software processing. The bright object near the top center is the 4.5 mag B5~V star HD~4727 ($\nu$~Andromeda). \label{fig1}
}
\end{center}
\end{figure*}

\begin{figure*}[ht]
\begin{center}
\includegraphics[angle=0,width=17.5cm]{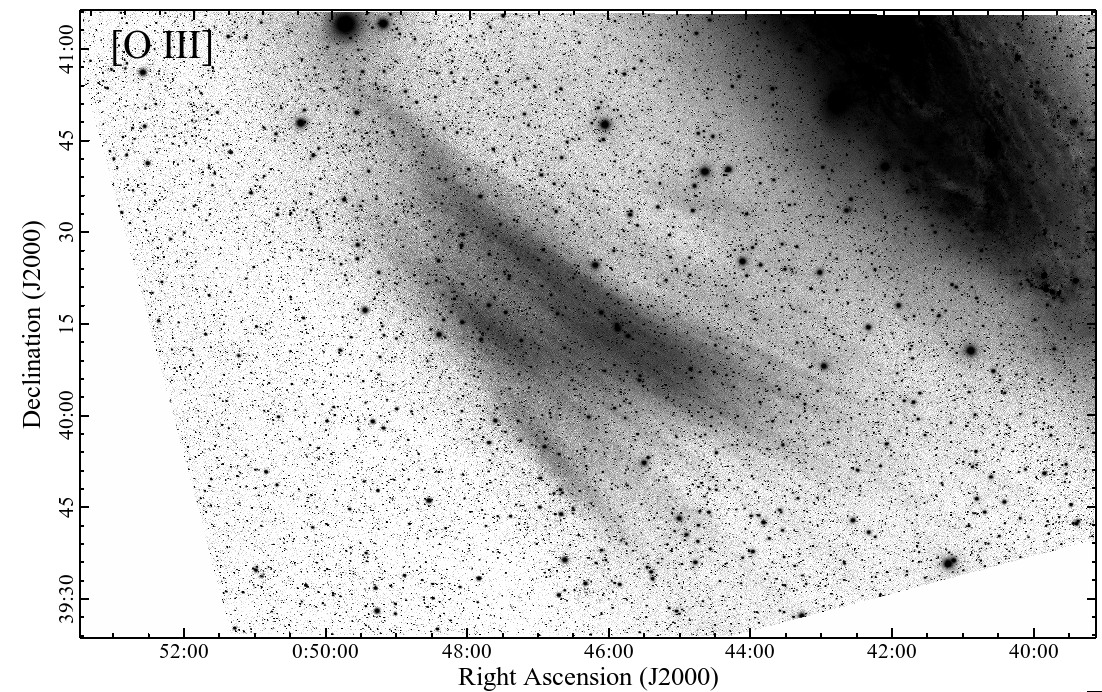} \\ 
\caption{Close-up of the $1.5\degr$ long [\ion{O}{3}] 5007 \AA \ emission nebula (SDSO) located southeast of M31's nucleus. \label{fig2} 
}
\end{center}
\end{figure*}

\begin{figure}[ht!]
 \centering 
  \includegraphics[width=\columnwidth]{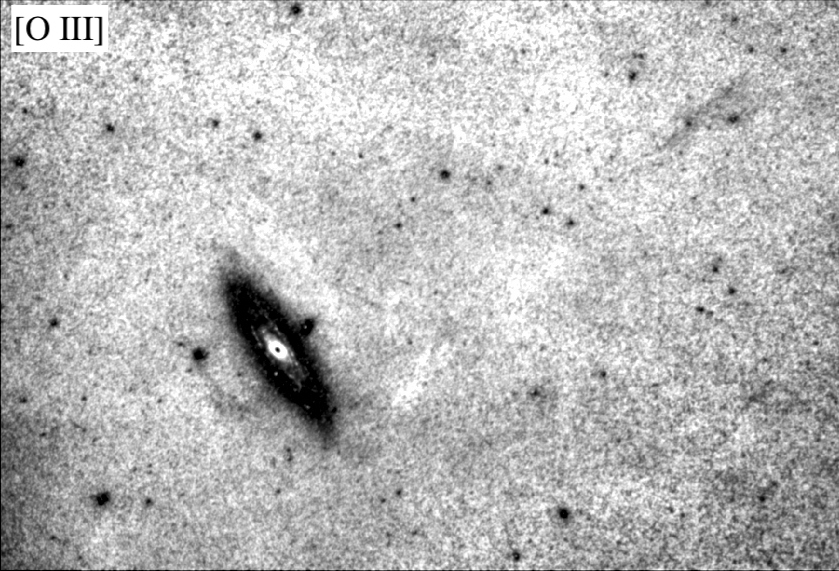}
\caption{Deep [\ion{O}{3}] image of the sky around M31. The FOV as shown is $9.2\degr \times 14.5\degr$.  North is up, East to the left.
\label{deep_O3}
}
\end{figure}

\begin{figure*}[t]
\begin{center}
\includegraphics[angle=0,width=18.8cm]{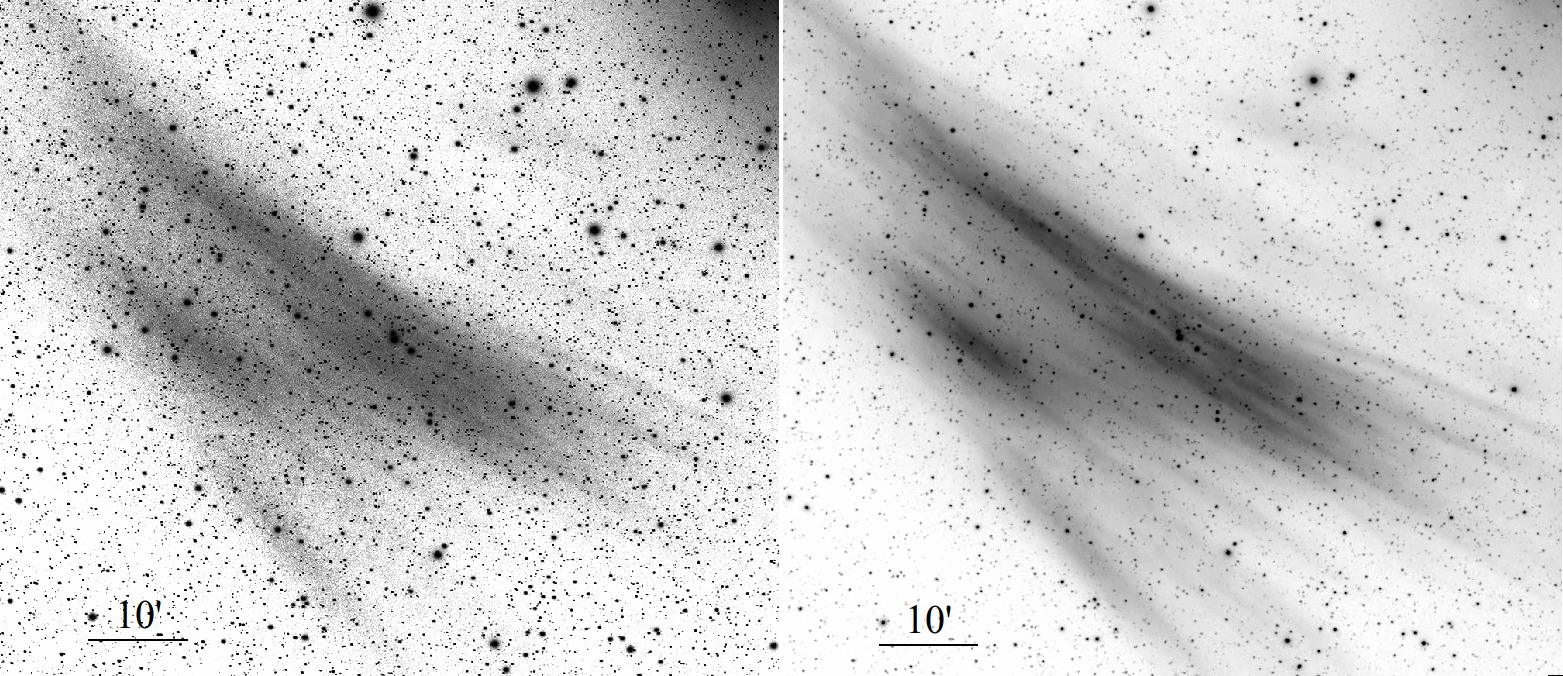} 
\caption{Spatial emission details of SDSO as seen in two independent sets
of deep [\ion{O}{3}] 5007 \AA\ images. Exposure totals were 46 hr and 110 hr for the left and right images, respectively.  North is up, East to the left.
\label{Yann_vs_Bray}
}
\end{center}
\end{figure*}

\section{Observations}

\subsection{Optical Imaging}

Images of M31 and its surroundings were obtained on 22 nights,
beginning in early 2022 August and continuing in early 2022 November
using [\ion{O}{3}] 5007 \AA, H$\alpha$ + [\ion{N}{2}]
$\lambda\lambda$6548,6583, and RGB continuum filters from several observing sites in France. These images were taken using a Takahashi FSQ106EDX4 telescope and a $6248 \times 4176$ pixel ZWO ASI2600MM Pro CMOS camera.
While this system provided a field of view (FOV) 
of $3.48\degr \times 2.32\degr$ with a 2.04$''$ pixel$^{-1}$ scale, shifts of pointing centers between images generated a slightly larger final FOV. 
Narrow passband interference filters (FWHM = 30 \AA) from Antlia\footnote{www.Antliafilter.com} were used. Total exposure times were 
45.7 hr ($274 \times 600$ s) in [\ion{O}{3}], 
41 hr in H$\alpha$ + [\ion{N}{2}] ($246 \times 600$ s), and 4 to 5 hr in each of the broadband RGB filters. 

A second set of independent images of M31 was obtained from September through November 2022 at a dark observing site in California\footnote{https://www.sierra-remote.com/} using two different telescope/camera systems. A series of [\ion{O}{3}] images totaling 85.5 hr ($513 \times 600$ s) plus 3.3 hr exposures in RGB filters were obtained using a Takahashi FSQ106mmEDX3 telescope and a $9576 \times 6388$ pixel QHY600 CMOS camera yielding a $5.33\degr \times 3.56\degr$ FOV. 
Narrow passband interference H$\alpha$ + [\ion{N}{2}]
$\lambda\lambda$6548,6583 (FWHM = 50 \AA) and [\ion{O}{3}]  (FWHM = 30 \AA) filters from Astrodon\footnote{https://astrodon.com/} 
with an image scale of 2.0$''$ pixel$^{-1}$ were used.
Additional [\ion{O}{3}] exposures totaling 24.9 hr ($299 \times 300$ s) using a Takahashi FSQ130mm telescope and a $9576 \times 6388$ pixel Moravian C3-61000 CMOS camera were also obtained, using a 30 \AA \ wide [\ion{O}{3}] filter from Chroma\footnote{https://www.chroma.com}.

Much wider FOV [\ion{O}{3}] images of M31 and the vicinity around it were obtained from various dark observing sites in France,
using a 135mm f/1.8 lens and a ZWO ASI2400MC Pro CMOS detector. This lens + camera  system gave a $15.3\degr \times 10.2\deg$ FOV with an image scale of $9.06''$ pixel$^{-1}$. A series of $84 \times 300$ s exposures totaling 7.0 hr was obtained.

\begin{figure}[ht!]
 \centering
  \includegraphics[width=\columnwidth]{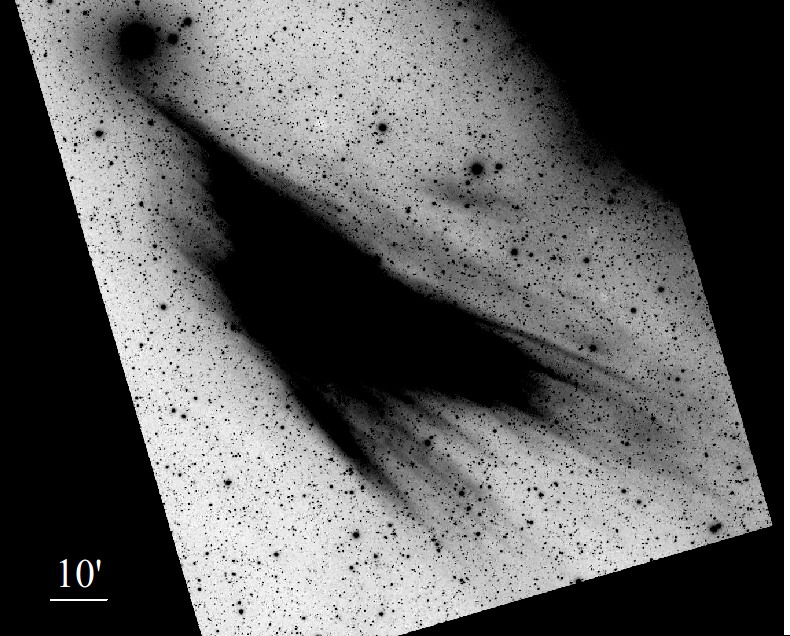}
\caption{High contrast [\ion{O}{3}] image showing SDSO's maximum extent. 
North is up, East to the left. 
\label{MAX_Bray} }
\end{figure}


\begin{figure}[ht]
\begin{centering}
\includegraphics[angle=0,width=8.6cm]{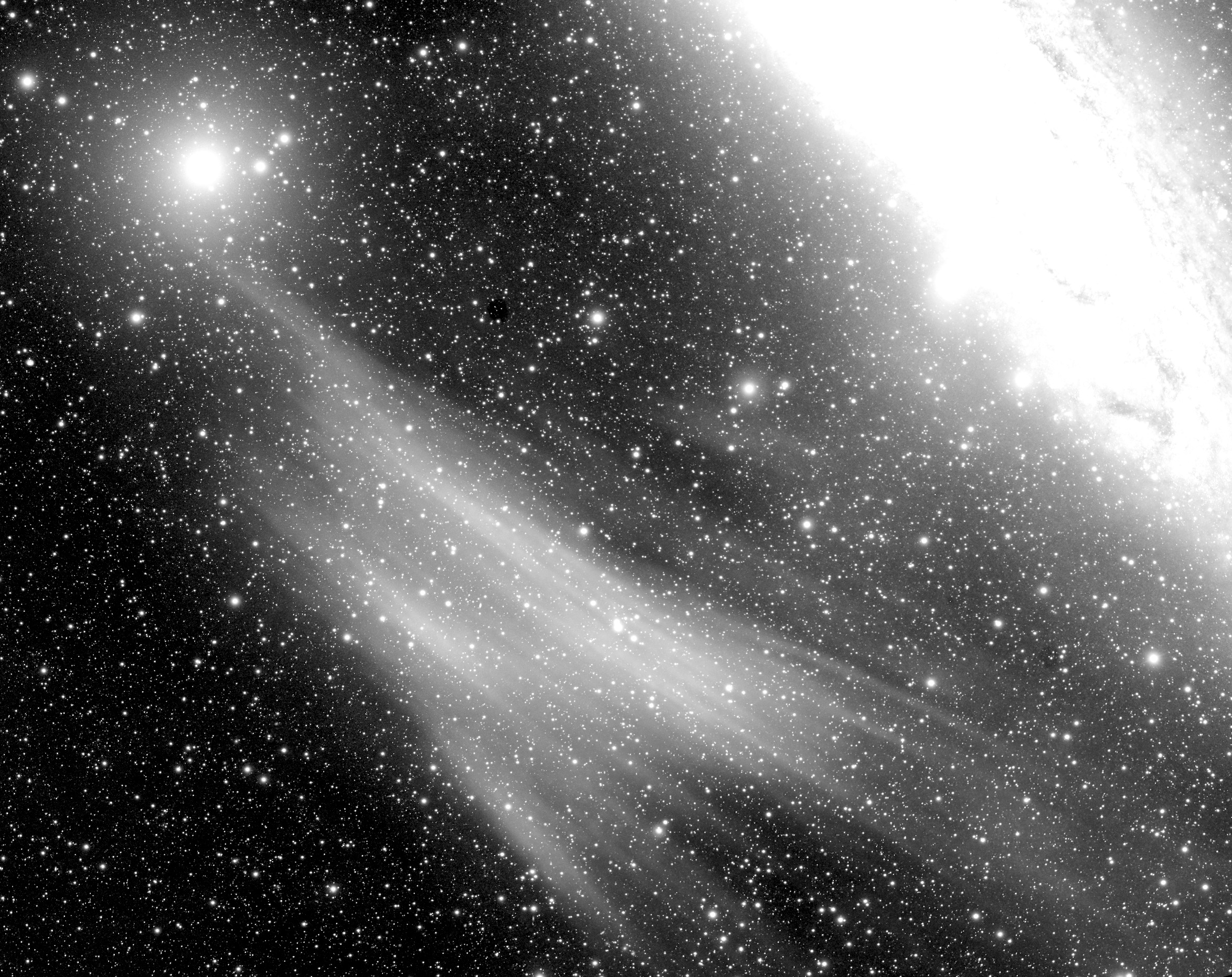} \\
\includegraphics[angle=0,width=8.6cm]{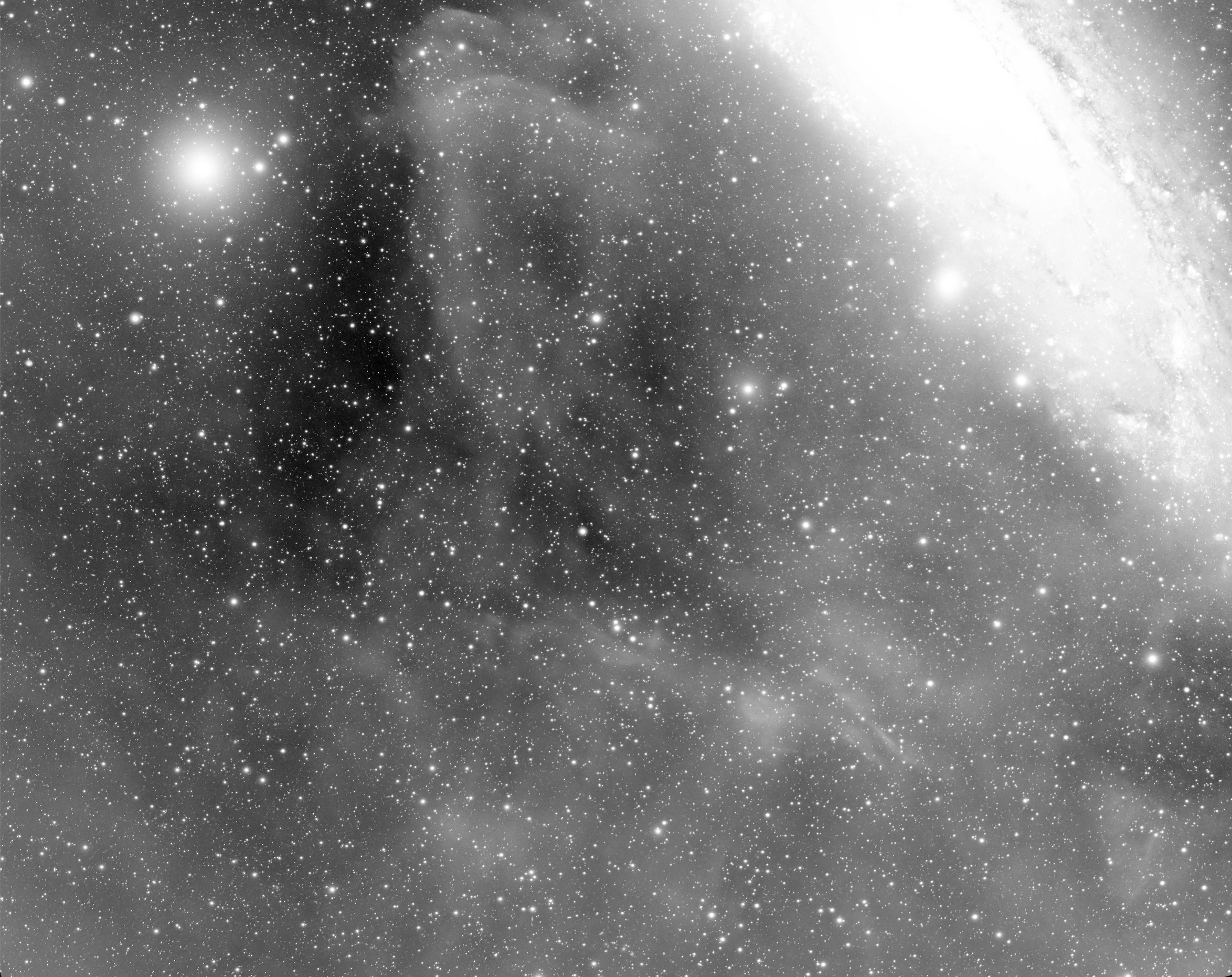}  \\
\includegraphics[angle=0,width=8.62cm]{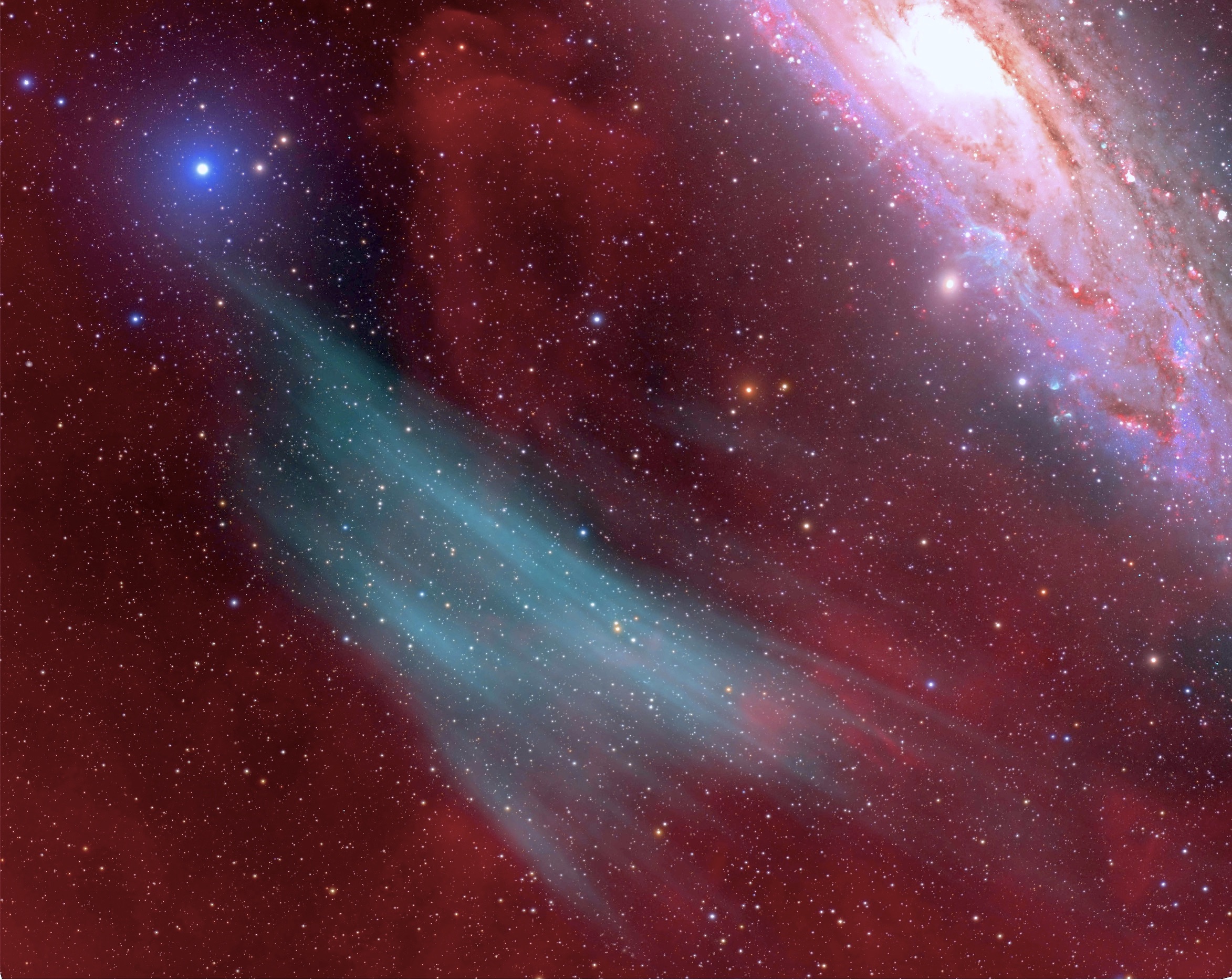}
\caption{Matching [\ion{O}{3}] (top) and H$\alpha$ + [\ion{N}{2}] (middle) images along with color composite (bottom) of SDSO 
and its immediate surroundings. North is up, East to the left.
 \label{fig6}
}
\end{centering}
\end{figure}

\begin{figure*}
\begin{center}
\includegraphics[angle=0,width=16.5cm]{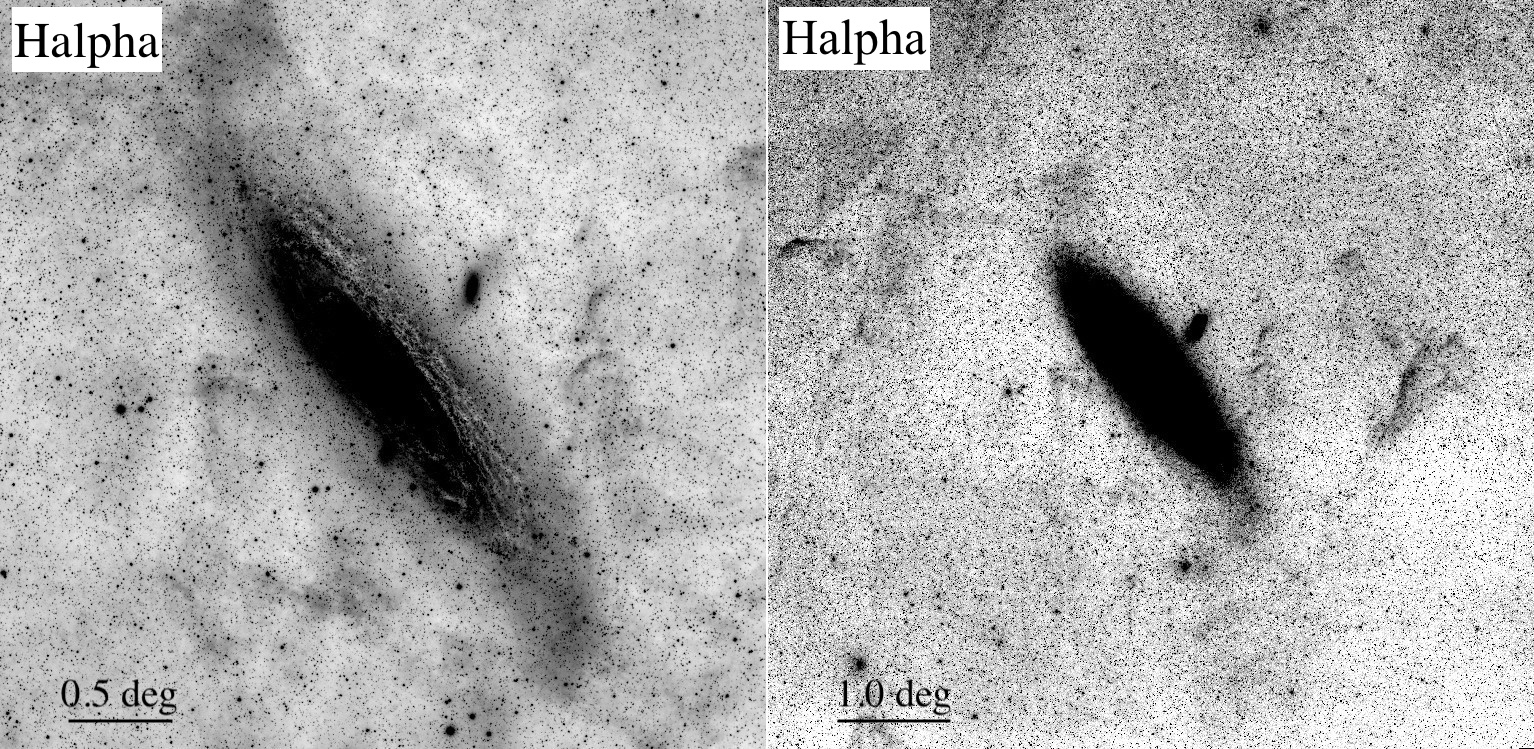} 
\includegraphics[angle=0,width=16.5cm]{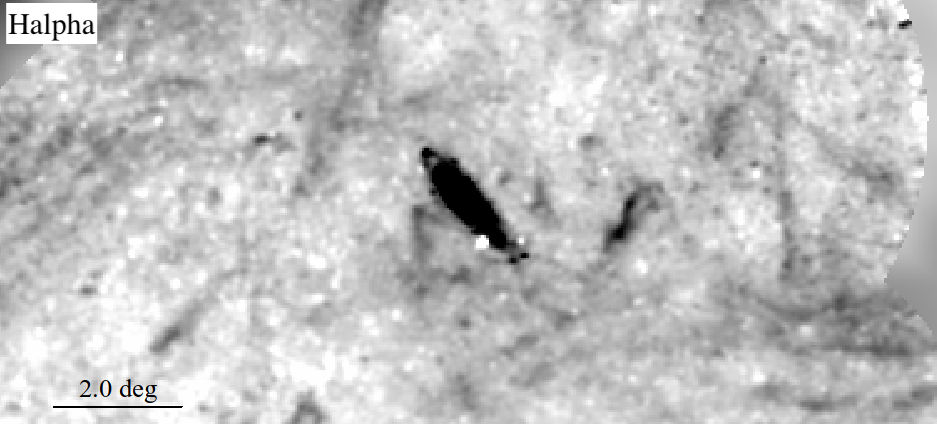}
\caption{
Wide field H$\alpha$ images showing faint emission around M31. 
Image FOVs as shown are: $3.8\degr \times 3.8\degr$ (top left) and $7.1\degr \times 7.1\degr$ (top right). Bottom: VTSS H$\alpha$ image;
$15.6\degr \times 7.1\degr$.  North is up, East to the left for all panels.
\label{fig7}
}
\end{center}
\end{figure*}

\begin{figure}[ht]
\begin{centering}
\includegraphics[angle=0,width=8.0cm]{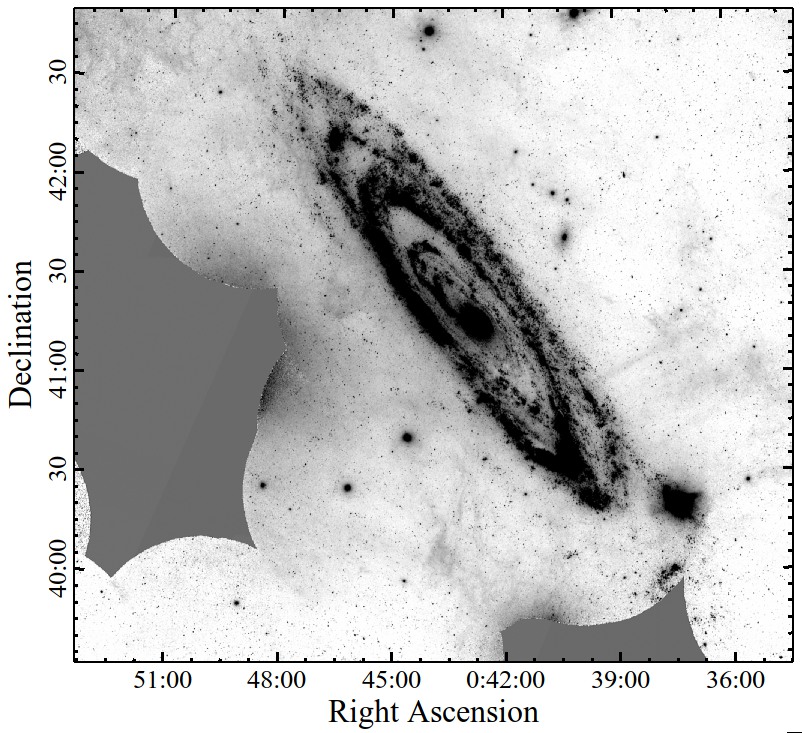} \\
\includegraphics[angle=0,width=8.00cm]{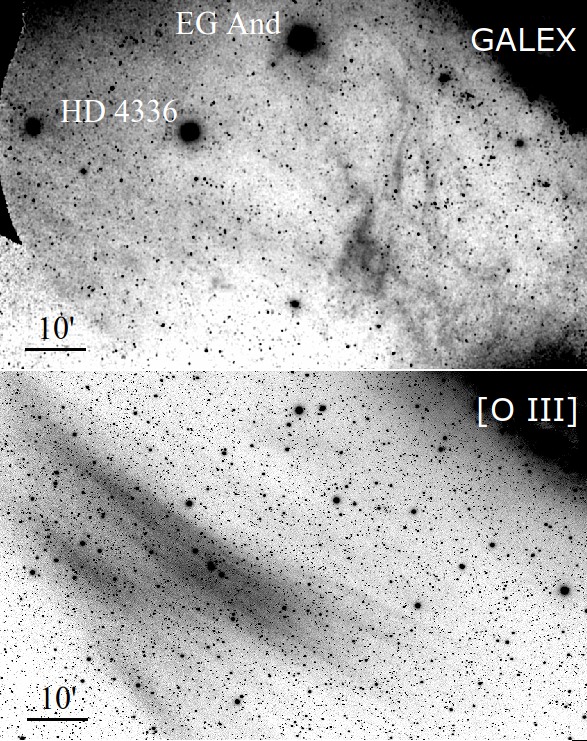} 
\caption{Top Panel: Mosaic of GALEX Far-UV (FUV) images of the M31 and its surroundings. Middle and Bottom Panels: Matching GALEX FUV (middle) and [\ion{O}{3}] (bottom) images of the SDSO region. The UV bright stars EG And and HD 4336 are marked in the GALEX image. 
 \label{fig8}
}
\end{centering}
\end{figure}

We also constructed a $9.2\degr \times 7.9\degr$ mosaic of H$\alpha$ + [\ion{N}{2}] image of M31 and the region around it from images taken as part of the MDW Sky Survey\footnote{https://www.mdwskysurvey.org}. This survey employs twin Astro-Physics 130~mm refractors, FLI Proline 16803 CCD cameras, and 30 \AA \ Astrodon H$\alpha$ filters. This imaging system has a FOV of $3.5\degr \times 3.5\degr$ with an image scale of $3.17''$ pixel$^{-1}$, with each region imaged for 4 hr ($12 \times 1200$ s).

Other deep H$\alpha$ images of M31 available in the public 
domain\footnote{https://apod.nasa.gov/apod/ap221024.html  \\
https://www.astrobin.com/b3f77y/?q=M3120H-alpha} and not part of our M31 imaging campaign were also examined.
One was obtained by A. Fryhover using a Rokinon 135~mm f/2.8 telescope and ZWO ASI2600mm pro CMOS camera. Exposures totaling 27.7 hr ($166 \times 600$ s) were obtained at a dark site in northwestern Oklahoma. Another deep H$\alpha$ image was obtained by V. Peris and A. Lozano from 51 hr of exposures 
taken at Observatorio Astronómico de Aras de los Olmos in Valencia Spain using a Canon EF400mm lens with an Astrondon 50 \AA \ filter and a QHY600L CMOS camera. This system yielded an image scale of $1.98''$ per pixel.  Finally, we examined the VTSS H$\alpha$ image of the M31 region \citep{Dennison1998}.

\subsection{Optical Spectra}

Low-dispersion optical spectra of the brightest regions of the arc were obtained with the MDM 2.4m Hiltner telescope using Ohio State Multi-Object Spectrograph (OSMOS; \citealt{Martini2011}). Spectra of SDSO were taken on 13 January 2023 UT (Slit 1) and again on 25 January 2023 UT (Slit 2). With a VPH grism (R $\simeq$ 1600) and a $3.0''$ wide north-south aligned slit, single exposures of 3600 s  each were taken at the two slit positions near the arc's north central edge
(Slit 1: RA(J2000) = 00:45:55.9; Slit 2: RA(J2000) = 00:45:39.8). The slit's effective $16'$ length allowed us to sample both the arc and the largely empty
region to the north (see Section 3.5 for further slit position details). 
While the nominal wavelength coverage for these spectra was 3400--5900 \AA, the camera/grism system was most sensitive in the 4000–5000 \AA \ region. For Slit 1, the spectra were binned along the slit ($0.55''$ pixel$^{-1}$). For Slit 2, we employed a $2 \times 2$ binning mode, with a dispersion scale 
1.45 \AA \ pixel$^{-1}$ and a FWHM resolution of 3.5~\AA. 

Additional spectra were taken at neighboring locations well off the arc for comparison purposes. For Slit~1, a 3600~s exposure was taken $0.65\degr$ to the southeast (J2000; RA = 00:48:32.85, Dec = +39:52:45) where no [\ion{O}{3}] is seen in our images. For Slit 2, a 3600~s exposure was taken immediately following the on-target exposure of a blank sky region (J2000; RA = 02:34:09.4, Dec = +40:19:45)
with similar hour angle and airmass values
to those of the on-target spectrum. For both slits, spectra were extracted after subtraction of off-arc spectra. Wavelength calibrations were made through Hg-Ne, Xe, and Ar comparison lamp spectra along with night sky lines. Data reductions were made using standard IRAF and MIDAS software routines.

\medskip

\subsection{GALEX UV Images}

We constructed wide-field GALEX UV images of the [\ion{O}{3}] emission region near M31 to investigate possible UV emission like that seen in the filamentary shocks of Galactic SNRs \citep{Fesen2021}. GALEX was a NASA science mission led by the California Institute of Technology, with 
a 50-cm diameter modified Ritchey-Chrétien telescope, a dichroic beam splitter and astigmatism corrector, and two microchannel plate detectors to simultaneously cover two wavelength bands with a $1.25\degr$ field of view with a resolution of $1.5''$ pixel$^{-1}$. 

Images were obtained in two broad bandpasses: a far-UV (FUV) channel sensitive to light in the 1344~\AA\ to 1786~\AA\/ range and a near-UV (NUV) channel covering 1771~\AA\ to 2831 \AA \ \citep{Morrissey2007}. The resulting images were circular with a FWHM resolution of $\sim4.2''$ and $\sim5.3''$ in the FUV and NUV bands, respectively.

\begin{figure*}[ht]
\begin{centering}
\includegraphics[angle=0,width=18.9cm]{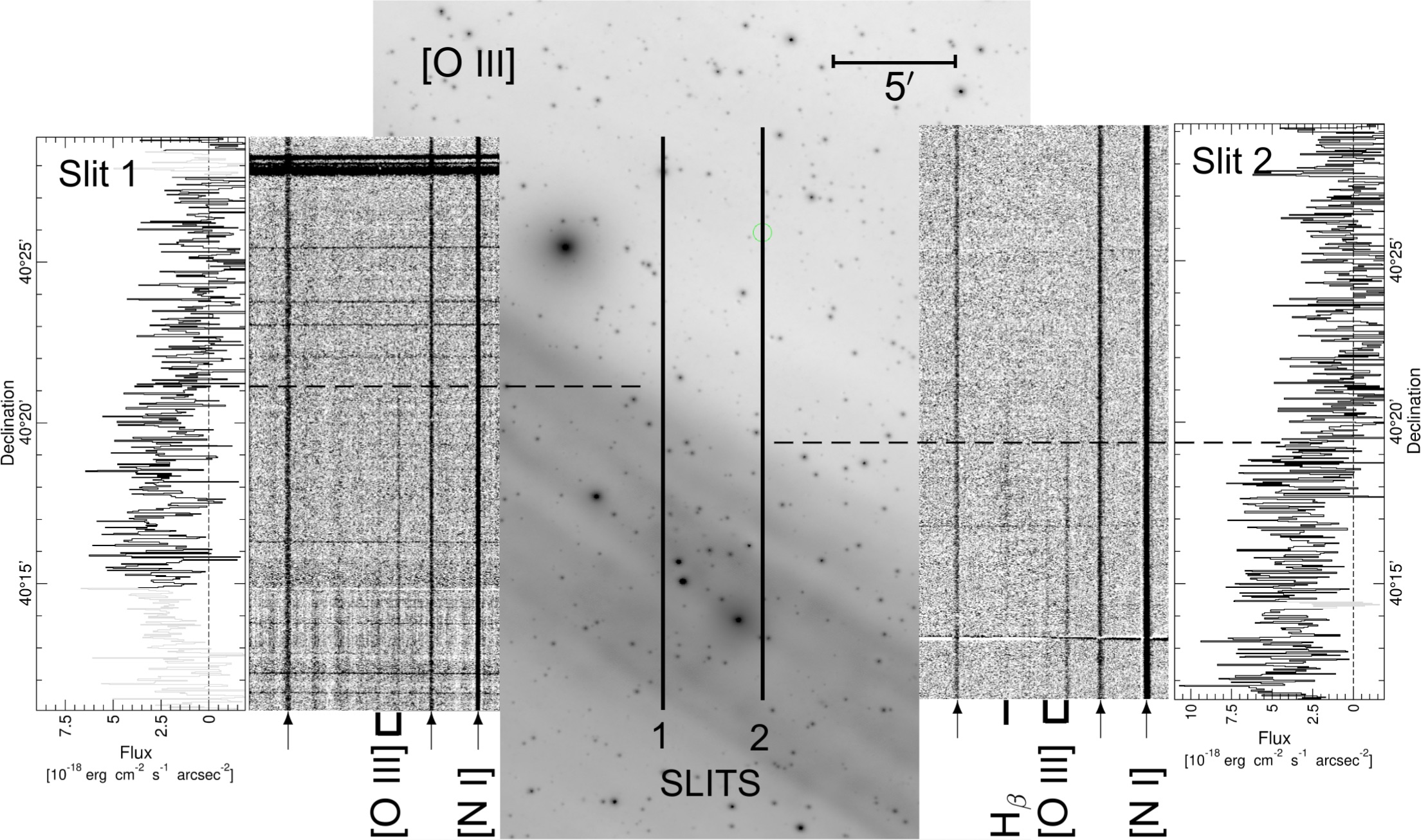}
\caption{SDSO Spectra: The middle panel shows our [\ion{O}{3}] $\lambda$5007 image of SDSO with the locations of Slits 1 and 2 indicated.
Adjacent left and right panels present resulting 2D spectra and the presence of [\ion{O}{3}] $\lambda\lambda$4959, 5007 emission,
which disappears, marked by dashed lines in the clear area north of SDSO. Night sky line emissions running the full length of the slits, including [\ion{N}{1}] $\lambda\lambda$5198, 5200, are marked by arrows at the bottom. The two outer panels show plots of the detected [\ion{O}{3}] $\lambda$5007 flux running from south to north along the lengths of both slits.
\label{fig9}
}
\end{centering}
\end{figure*}

\begin{figure}[ht]
\begin{centering}
\includegraphics[angle=0,width=8.9cm]{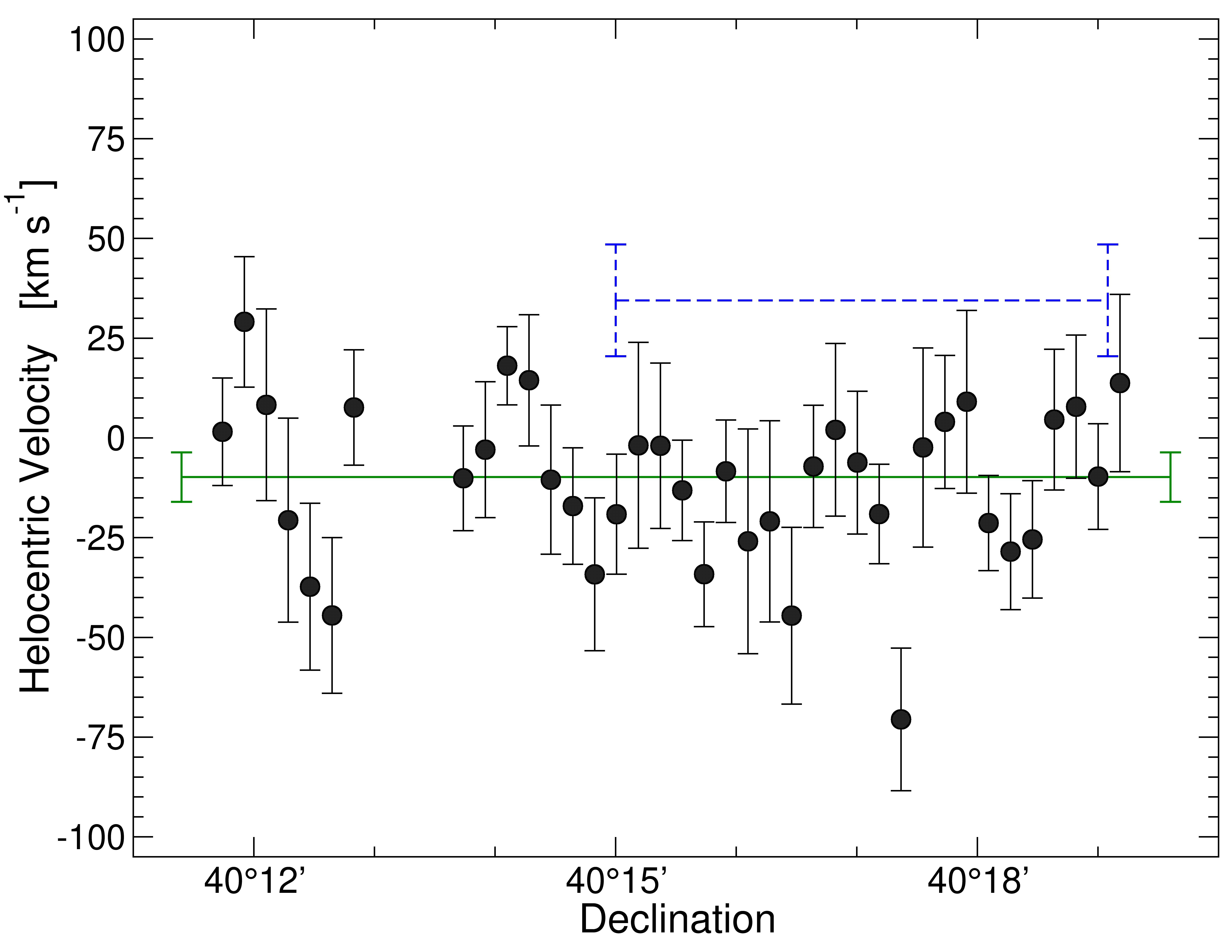} 
\caption{Plot of heliocentric velocities for [\ion{O}{3}] $\lambda$5007 (green line) and H$\beta$ (dashed blue line) along the Slit 2 moving from south to north. The error bars indicate the statistical errors for the global fit solution,
$-9.8 \pm 6.8$ km~s$^{-1}$.
 \label{fig10}
}
\end{centering}
\end{figure}

\begin{figure}[ht]
\begin{centering}
\includegraphics[angle=0,width=8.7cm]{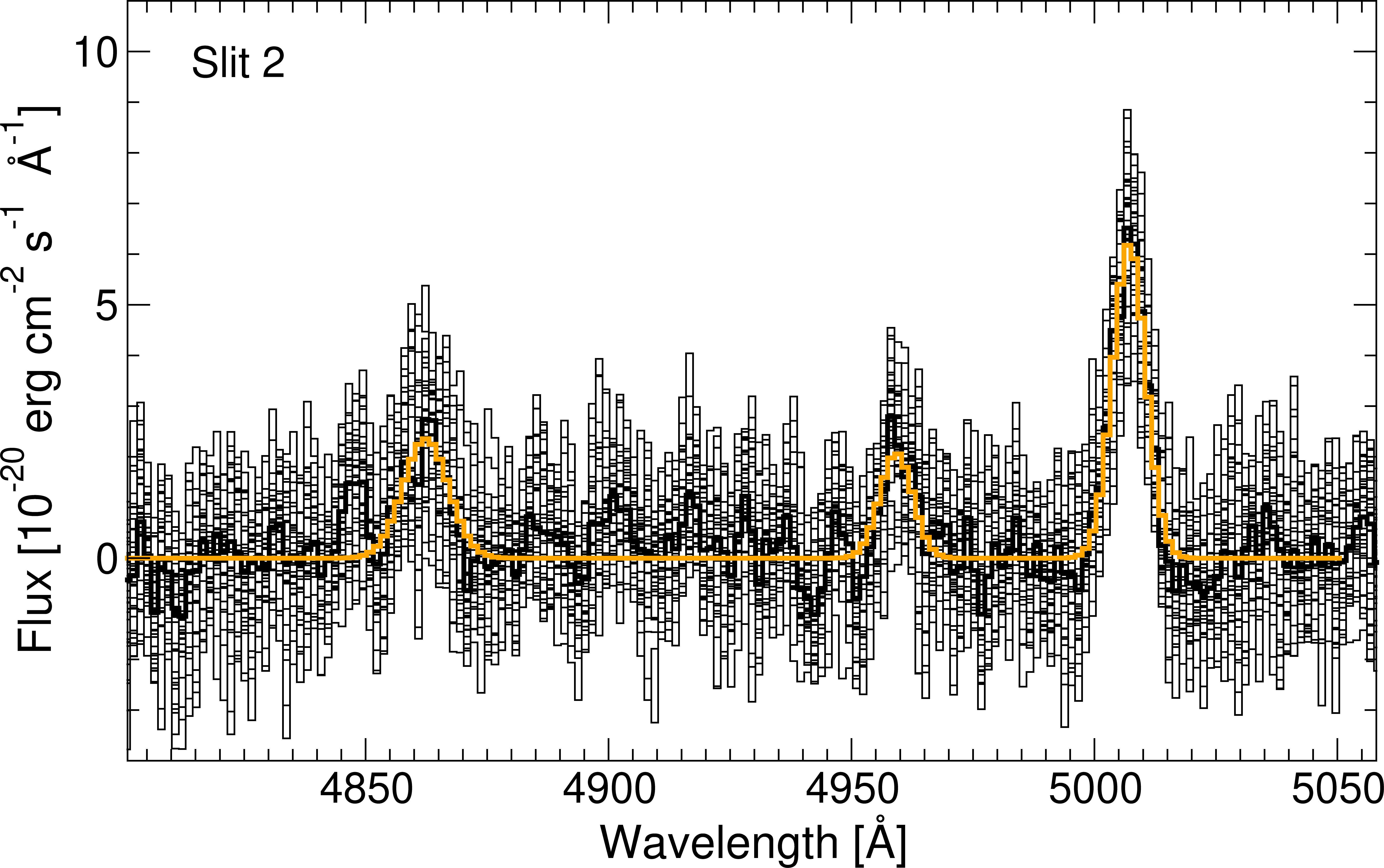} 
\caption{Spectra obtained in Slit 2 around H$\beta$ and [\ion{O}{3}] integrated in $10''$ bins in the region of SDSO between $+40\degr 11' 25''$ $< \delta <$ 
$+40\degr 19' 10''$ (gray lines) and the total averaged spectrum (orange line). The wavelengths shown are in the observed telluric rest frame. The [\ion{O}{3}] $\lambda$4959 line fit used that measured for the
[\ion{O}{3}] $\lambda$5007 line.
 \label{fig11}
}
\end{centering}
\end{figure}


\section{Results}

Figure 1 shows the original sum of 45.7 hr of [\ion{O}{3}] images, revealing an emission nebula near M31 discovered by \citet{Drechsler2023} and
hereafter referred to as either SDSO or simply as the M31 [\ion{O}{3}] arc. The left panel shows the faintness of the arc, even using a narrow passband 30 \AA\ filter and illustrates why the arc was missed in wide band images. Subtraction of background blue and green continuum images makes the nebula obvious, as seen in the negative version in the righthand panel. 

The arc's small-scale details and coordinates can be seen in the enlarged 
Figure 2.  Estimated to be $\simeq1.5\degr$ in length, $\simeq0.5\degr$ in width and centered some $1.2\degr$ southeast of M31's nucleus \citep{Drechsler2023}, this line-emission nebula appears composed of several broad and gently curved filaments aligned approximately NE-SW. Although displaying a filamentary appearance, the nebula's overall structure is mainly diffuse. No thin or sharp filaments are visible, as often seen in [\ion{O}{3}] SNR filaments and in some PNe.

The nebula is composed of three main and largely aligned emission regions. The nebula's brightest region lies closest to M31 and extends farthest to both the northeast and southwest. In Figure~2, none of the emission regions exhibit well-defined ends, but instead gradually fade at their NE and SW ends.  While the nebula is fairly well defined along its eastern portions, along its western edges, fainter emission bands and patches can be seen lying in between it and M31's disk.

\subsection{No Other [\ion{O}{3}] Emission Nebulae Near M31}

\citet{Drechsler2023} presented [\ion{O}{3}] images of M31's eastern region,
extending $2\degr$ from the M31 nucleus and to the west past NGC 205. These data revealed no [\ion{O}{3}] emission nebulae around M31 other than SDSO.
To investigate this further, we obtained wider FOV [\ion{O}{3}] images. These indicate that, with the exception of an emission cloud $8\degr$ northwest of M31, SDSO appears to be an isolated [\ion{O}{3}] nebula located relatively close in projection to M31. 

This can be seen in Figure 3, where we present a $9.2\degr \times 14.5\degr$   [\ion{O}{3}] image of M31 and its surroundings.  The curvature of SDSO 
and the lack of any detected [\ion{O}{3}] emission outside of M31's disk on the opposite side of M31 from SDSO suggest that this is an isolated emission feature and not part of a larger [\ion{O}{3}] emission structure.
Because this image is the result of only 7 hr of exposure, SDSO is more poorly detected in this image compared to our other, deeper images. Consequently, we cannot rule out the presence of other extended [\ion{O}{3]}] nebulae around M31 with a surface brightness much lower than that of SDSO. Nonetheless, it is clear that no similarly bright [\ion{O}{3}] emission feature lies within several degrees of M31.

\subsection{SDSO's Substructure and Extent}

An independent and slightly deeper [\ion{O}{3}] image set confirms and clarifies the nebula's fine-scale structure. This is shown in Figure~4, 
where the left panel shows the discovery [\ion{O}{3}] image made from 45.7 hr of exposures \citep{Drechsler2023}, with the right panel image composed of a separate set of 110 hr of [\ion{O}{3}] exposures. Both images as shown are the products of significant post-processing using various commercial image software. However, the fact that nearly identical arc substructures are seen in the two images obtained using different telescope + camera systems supports the reality of SDSO's substructures as seen in these images. 

We note that since the images were taken over a long period of time spanning months, telluric airglow emissions and other night sky contamination can be ruled out. Small scale structures change on timescales of minutes and at most hours \citep{Noll2012,Jones2019} and only add a constant background.  Seasonal variations of the background causing long time-scale biannual fluctuations are well studied  \citep{Noll2017} and can also be ruled out. Finally, post processing was done on individual images near in time blocks before co-adding with all image subgroups show the same morphology.

The longer exposure image (Fig.\ 4, right panel) shows SDSO to be entirely diffuse in structure, with its filamentary appearance mainly due to several long and broad streaks of diffuse emission rather than thin, sharp filaments. This longer exposure image also shows that SDSO's emission extends farther to the SW and off the image frame shown. 

The full extent of SDSO's emission can be seen in the higher contrast version presented in Figure 5.  
This shows that the nebula's maximum NE-SW length is larger than the $1.5\degr$ estimated by \citet{Drechsler2023} and is closer to $2\degr$, extending from its NE tip near the B5~V star HD~4727 ($\nu$ And) down to the far right-hand corner (RA = 00:41, Dec = +39:45)
where the emission fans out into two broad diffuse branches.  
Lastly, this higher contrast image also highlights the arc's nearly straight eastern edge. Interestingly, this eastern edge exhibits a position angle (PA) of $\simeq$ 40$\degr$ which is nearly the same as the PA of $38\degr$ for M31's major axis \citep{deV1958,Walterbos1987}.  

\vspace{1cm} 

\subsection{Projected H$\alpha$ Emission Nebulae Near M31}

\citet{Drechsler2023} reported finding no bright H$\alpha$ emission 
nebulosity coincident with the SDSO nebula and estimated a ratio
I([\ion{O}{3}]$\lambda\lambda$4959,5007)/I(H$\alpha$)
$\geq5$ for the nebula's brightest regions based on deep but uncalibrated [\ion{O}{3}] and H$\alpha$ images. Below we discuss deep H$\alpha$ images near M31, obtained both by ourselves and others, which indicate no strong coincident H$\alpha$ emission at SDSO's location.

Figure 6 shows positive [\ion{O}{3}] and H$\alpha$ + [\ion{N}{2}] images along with a color composite of these images.  This figure shows that there is little  H$\alpha$ emission coincident with the [\ion{O}{3}] arc. The detected H$\alpha$ emission displays a very different morphology to that seen in the [\ion{O}{3}] image suggesting no association with the arc. This is highlighted in the color image which suggests a high 
[\ion{O}{3}]/H$\alpha$ ratio for a majority of the arc. 

Figure 7 presents wider FOV H$\alpha$ + [\ion{N}{2}] images of M31 and 
its immediate neighborhood obtained with a variety instruments and exposures\footnote{Note: The top images labeled ``H$\alpha$'' 
were taken with filters also sensitive to the usually much weaker [\ion{N}{2}]
$\lambda\lambda$6548,6583 emission lines. The bottom VTSS image was taken using a 17.5 \AA \ wide filter that avoided most [\ion{N}{2}] emission contamination.}. The top right panel image was obtained by us, and the top left was obtained by
A. Fryhover. (Note: A nearly identical H$\alpha$ image to Fryhover's image was obtained by V.\ Peris and A.\ Lozano\footnote{https://www.astrobin.com/b3f77y/}.)
The bottom panel shows a much wider FOV image taken as part of the Virginia Tech Spectral-line Survey (VTSS). 

These H$\alpha$ images and those shown in Figure~7 look remarkably similar, despite the range of H$\alpha$ filter widths used; namely, 50 \AA\ for the top right image, 30 \AA\ for the top left image, and 17.5 \AA \ for the VTSS image (bottom panel). Since SDSO was detected using a 30 \AA \ wide [\ion{O}{3}] $\lambda$5007 filter, the lack of correlated H$\alpha$ emission with SDSO does not appear to arise from radial velocity differences between emitted [\ion{O}{3}] and H$\alpha$ line emissions

Most of the detected diffuse H$\alpha$ emission seen here is likely Galactic emission unrelated to M31 with the exception of extended diffuse emission from M31's warped disk immediately north and south of M31's bright main disk.
Outside of the different FOVs and image resolutions, we find no significant differences in the detected emission features among these H$\alpha$ images. Most importantly, no significant H$\alpha$ emission is seen in the location of SDSO, and none show a similar structure or extent.

\subsection{GALEX Images of M31 and the SDSO Region}

Nebulae exhibiting bright [\ion{O}{3}] emissions are sometimes accompanied by significant far UV line emissions, such as seen in SNRs (\citealt{Kim2014,Tutone2021,Fesen2021}).
Because of SDSO's strong [\ion{O}{3}] emission, we have examined GALEX FUV images of M31 to see if SDSO also exhibits detectable FUV emission.
A mosaic of GALEX FUV images of M31 and its surroundings is shown in Figure 8.
Although wide field GALEX images of M31 have been presented before \citep{Thilker2005b}, most subsequent UV studies concentrated on M31 point sources and star clusters \cite{Kang2009,Kang2012,Bianchi2014,Leahy2021}.
None of these papers commented on extended emission outside of the M31 disk.

Figure 8 shows no obvious correlated FUV emission at SDSO's location. 
However, given the extreme faintness of SDSO in [\ion{O}{3}], it is not clear that a lack of FUV emission seen by GALEX is especially meaningful.
We do find considerable diffuse and filamentary nebulosity, seeming to extend radially outward from M31's southern disk, partially coincident with M31's Giant Stellar Stream of stars (GSS; \citealt{Ibata2001,McConn2003,McConn2005,vanderMarel2012,Gilbert2007,Gilbert2019,
Fardal2007,Fardal2012}).
To our knowledge, the presence of far UV filaments at the location of M31's GSS has not been previously reported despite wide FOV mosaic GALEX FUV images covering this region having been published \citep{Thilker2005b,Madore2005}.

\subsection{Optical Spectra}

Low-dispersion optical spectra were obtained at two locations near the northern limb of SDSO. The slit positions, labeled Slits 1 and 2, are shown in the middle panel of Figure 9.  The slits are $16'$ in length and cover a portion of SDSO as well as an [\ion{O}{3}] emission-free region to the north. 

The resulting long-slit 2D spectra obtained at both slit positions are shown in adjacent panels of Figure 9. These show a clear detection of [\ion{O}{3}] emission in both slit spectra corresponding to the location of SDSO. Its disappearance along the slits is marked by the horizontal dashed lines at the upper edge of the detected emission, as seen in the [\ion{O}{3}] image. 
In the brightest regions, faint emission from both [\ion{O}{3}] $\lambda\lambda$4959,5007 lines can be seen. Unlike the detected [\ion{O}{3}] line emissions, several night sky lines, marked by arrows at the bottom, are seen to be continuous along the full length of the slits.

The outermost panels in Figure 9 show the detected flux for the
[\ion{O}{3}] $\lambda$5007 line.  The flux drops sharply at the northern 
edge of SDSO, matching our [\ion{O}{3}] image. No [\ion{O}{3}] line emission was detected with identical exposure times in a region $40'$ to the SE from SDSO.  Because [\ion{O}{3}] emission is seen in both Slits 1 and 2 and fades from south to north along the slit, we are confident that we have detected [\ion{O}{3}] line emission principally from SDSO.

Using night-sky telluric emission lines as zero-point calibrators, notably the
[\ion{N}{1}] doublet lines at 5197.90~\AA\ and 5200.26~\AA,
we find heliocentric velocities of $-9.8 \pm 6.8$ km~s$^{-1}$ for the 
[\ion{O}{3}] emission lines and $+34 \pm 14$ km~s$^{-1}$ for the brighter portion of the detected H$\beta$ emission seen in the spectrum taken at 
Slit~2.  The individual [\ion{O}{3}] velocity measurements taken at $10''$ increments along Slit 2 are shown in Figure 10, along with the blue line representing the average velocity for the H$\beta$ line.

Figure 11 shows the integrated spectrum for Slit 2 with the wavelengths in the observed telluric rest frame.
The displayed Gaussian fit for the [\ion{O}{3}] $\lambda$4959 line used the [\ion{O}{3}] $\lambda$5007 line  scaled down 1:3 and shifted to the same radial velocity of $-9.8$ km~s$^{-1}$ of the \ion{O}{3}] $\lambda$5007 line. 
The H$\beta$ fit results in a wider line with a radial velocity of +34 km~s$^{-1}$.
Our velocity results indicate that the H$\beta$ emission seen along part of Slit 2's spectrum is not related to SDSO's [\ion{O}{3}] emission. This is consistent with our images, which show little H$\alpha$ at SDSO. 

The surface brightness of SDSO's brighter regions estimated from the data ranges from $(2.5\pm2) \times 10^{-18}$ erg s$^{-1}$ cm$^{-2}$ arcsec$^{-2}$ for Slit 1, to $(4\pm2) \times 10^{-18}$ erg s$^{-1}$ cm$^{-2}$ arcsec$^{-2}$ for Slit 2. These values are consistent with our earlier estimate reported in \citet{Drechsler2023}. These values drop sharply to under
$1 \times 10^{-18}$ erg s$^{-1}$ cm$^{-2}$ arcsec$^{-2}$ at SDSO's northern edge in this region.  Using the conversion of 
1 Rayleigh = $7.42\times 10^{-18}$ erg s$^{-1}$ cm$^{-2}$ arcsec$^{-2}$ for [\ion{O}{3}] $\lambda$5007, we find that SDSO's brighter regions have a surface brightness in [\ion{O}{3}] of $\sim 0.55$ Rayleigh, and the faintest regions are detected at $\sim 0.2$ Rayleigh.

We note that our estimated SDSO's [\ion{O}{3}] flux and velocity are noticeably different from those reported by \citet{Amram2023} who also obtained spectra of SDSO in January 2023 but at a lower resolution than our data. They cite an [\ion{O}{3}] velocity of $-96\pm4$ km~s$^{-1}$, although their listed accuracy seems inconsistent with the resolving power $R \approx 750$ (400~km~s$^{-1}$) of their low-resolution spectrograph.
They also report detecting strong H$\alpha$ and [\ion{N}{2}] emissions and an [\ion{O}{3}] surface brightness seven times brighter than our estimate.  They cite a surface brightness value of $(2.7\pm1.4) \times 10^{-17}$ erg s$^{-1}$ cm$^{-2}$ 
arcsec$^{-2}$ (3.7 Rayleigh for [\ion{O}{3}] $\lambda$5007).

The source of the discrepancy between their and our measurements is not obvious.  It may simply be that their slit sampled a spot a few arc minutes south of our two slit positions where there is a bright diffuse clump of H$\alpha$ emission (see our Fig.\ 6). This could explain their strong H$\alpha$ and [\ion{N}{2}] emissions. However, whereas our 2D spectra clearly show detected [\ion{O}{3}] from SDSO, [\ion{O}{3}] emission is not visible in their 2D background subtracted spectra (their Fig.\ A.1). In addition, they found H$\alpha$ emission to be roughly twice as bright than [\ion{O}{3}] 
(i.e., H$\alpha$/[\ion{O}{3}] 5007 = 1.5 - 2.8), whereas our images of SDSO show no strong or correlated H$\alpha$ emission (see Fig.~6).

\section{Discussion}

The morphology and emission properties of SDSO do not lead us, by themselves, to a definite conclusion about its location and physical nature.
The fact that it is presently only detected in optical [\ion{O}{3}] line emission, together with the lack of a constraint on its distance and hence its physical size, severely limits our ability to determine its likely nature.

The goal of our spectra of the emission arc was to provide radial velocity information which might establish (or not) an association with M31 and its halo. The Local Group, with M31 and the Milky Way as its two main components, is a bound system decoupled from the Hubble expansion.
M31's heliocentric velocity is $-300 \pm 4$ km s$^{-1}$ \citep{Slipher1913,deVau1991,McConn2005}
of which $\sim 200$ km s$^{-1}$ is due to the directional component of the Sun's
velocity around the Galaxy's center.  This leaves the M31 barycenter with an approaching velocity to the Milky Way barycenter of $\sim$110 km s$^{-1}$.

Our estimated heliocentric radial velocity for SDSO of roughly $-10$ km s$^{-1}$ 
would seem, on its face, to strongly favor a Galactic origin, rather than 
one associated with M31.  However, for reasons discussed below, we view a Galactic origin as less likely than an extragalactic origin associated with M31.
Below we briefly discuss various possibilities as to the nature of SDSO, divided into Galactic and M31 scenarios. We pose three Galactic possibilities: an undiscovered SNR at an unusually high Galactic latitude, a faint and unusually nearby planetary nebula, or a stellar bow shock nebula. We then discuss the possibility that SDSO is an extragalactic nebula possibly related to M31 and offer explanations for its small heliocentric radial velocity. 

\subsection{A Galactic Origin}

\subsubsection{A Faint, Undiscovered Galactic SNR}

A large and curved nebula displaying strong [\ion{O}{3}] emission relative to that of H$\alpha$ might indicate the presence of a faint and previously unknown SNR with a location that just happens to be near M31 by chance.
Indeed, based on low-dispersion optical spectra, \citet{Amram2023} tentatively proposed that SDSO is the most visible part of a large 35-pc diameter SNR projected near M31 located some 0.7 kpc away. That would correspond to a 0.05-radian ($2.86\degr$) angular diameter, placing the far side of the remnant $\sim1.7\degr$ northwest of the M31 nucleus, where no emission is apparent (see Figs.\ 3 and 7).  

At a Galactic latitude  $b = -22.5\degr$, a very low surface brightness supernova remnant located near M31 in the sky might have escaped notice in Galactic SNR radio surveys, which have concentrated on regions within $\approx 5\degr$ of the Galactic plane (e.g., \citealt{Reich1990,Duncan1995,Langston2000}). 
SDSO might also be one of those SNRs where some portion of their optical structure is dominated by [\ion{O}{3}] line emission. 

The most cited case for this situation is that of the $5.2\degr \times 4.0\degr$ remnant G65.3+5.7 located nearly six degrees above the Galactic plane \citep{Gull1977}. This remnant was only discovered due to its strong [\ion{O}{3}] emission in the \citet{Parker1979} survey.  Nearly all of this remnant's optical emission structure consists of high I($\lambda$5007)/I(H$\alpha$) ratios with values as large as $\simeq10$ \citep{Fesen1985,Mav2002,Boumis2004}. While such line ratios are unusual, some remnants display strong [\ion{O}{3}] emission over small portions of their optical structure.  These include the Cygnus Loop \citep{Fesen1996}, CTB~1 \citep{Hailey1994,Fesen1997}, and G179.0+2.6 \citep{How2018}.

If SDSO were part of an undiscovered SNR, its slight curvature would suggest a fairly large remnant spanning many degrees in angular size. Judging by the arc's most strongly curved inner feature, we estimate a radius $\sim2\degr$ assuming a perfectly circular structure. A Galactic remnant this size would be unusual but hardly extraordinary, and its center would lie to the northwest of the M31 center. 
A SNR located some 22 degrees off the Galactic plane would not likely have a vertical distance more than $z \sim$1.5~kpc. Consequently, an angular diameter 
of $4\degr$ would imply physical dimensions of 35--105~pc if located 
0.5--1.5~kpc distant. These dimensions are typical for a middle-aged or an old SNR. SDSO's strong [\ion{O}{3}] emission relative to H$\alpha$ would suggest a shock velocity $\simeq$ $90 - 150$ km s$^{-1}$.  This is a common range among young to middle-aged SNRs with strong [\ion{O}{3}] emission, but unlikely for an old remnant $\sim$ 100 pc in diameter. Consequently, if SDSO were a SNR it would likely be of moderate age and among the closest SN remnants known.

While a SNR nature for SDSO is possible, there are difficulties with a SNR explanation, primarily the lack of any other portion of a supposed remnant shell. As shown in Figure 3, SDSO is the only [\ion{O}{3}] nebulosity around M31. In addition, we found no hint of a possible associated H$\alpha$ emission shell around M31 that might complete a SNR shell.  In that scenario, SDSO might be just one part that is especially bright in [\ion{O}{3}] line emission. If it was a large, nearby SNR located well above the Galactic plane with low expected extinction, the lack of any H$\alpha$ or additional [\ion{O}{3}] emission from other portions of a SNR shell would be quite unusual.

Moreover, although strong [\ion{O}{3}] emission in SNRs is not uncommon, this type of SNR emission almost always has the morphology of sharp, thin filaments marking the remnant's outermost shock boundary. This is not what is seen. Instead, SDSO exhibits a broad structure with little evidence of limb brightening, as would be expected if it were part of a large SNR emission shell. In addition, the individual features that make up SDSO do not all display the same degree of curvature.  The sections closest to M31 are the most curved, with the more eastern region almost straight, with little if any curvature.

The fact that SDSO's emission faces away from the Galactic plane is also opposite from that expected if the observed emission was due to an interaction of a 
high-latitude remnant's expanding shell with interstellar clouds or into the ISM's density gradient above the disk plane. Although there are examples where a high latitude remnant's optical emission is located farthest away from the Galactic plane (e.g., G116.6-26.1; \citealt{Pal2022}), it is more common to find high latitude SNRs with their brightest optical emission concentrated closest to the Galactic plane (e.g., G70.0-21.5, G74.0-8.5 (Cygnus Loop), G107.0+9.0, G249.7+24.7, G275.5+18.4 (Antlia),
and G354.0-33.5; \citealt{Fesen2015,Fesen2018,Fesen2020,Fesen2021}).

Another problem is the lack of any detected nonthermal radio emission at SDSO's location. Despite numerous radio surveys of M31 and local environs, 
there is no reported nonthermal emission in the SDSO area southeast of M31's nucleus.  In both low and high frequency radio and far infrared surveys of M31, there has been no reported adjacent SNR or any evidence of emission in M31's halo at SDSO's location so close to M31's disk
\citep{Graeve1981,Beck1982,Berk1983,Berk1991,Beck1998,Berk2003,Planck2015,Harper2023}. While SDSO's faint optical emission might suggest equally faint radio emission, the lack of any hint of coincident radio emission is a potential difficulty. 

We also note the small chance of discovering a SNR at Galactic latitude 
$b = -22.5\degr$ plus one so close to M31 in projection.  Although several large SNRs have recently been discovered located at similar and even greater distances above the Galactic plane \citep{Churazov2021,Becker2021,Fesen2021,Pal2022}, only six currently known SNRs have Galactic latitudes $|b| \geq 15\degr$, representing less than 2\%  of the 306 currently known Galactic SNRs \citep{Green2022}. This rarity, combined with a projected location within just 1.5 degree of M31 and having its NE-SW structure in near alignment with M31's disk, makes the random chance of SDSO being a high latitude SNR near M31 appear quite small ($\lesssim 10^{-2}$).

Although coincidental alignments have occasionally 
been seen for supernova remnants\footnote{For example, a compact flat spectrum radio source at the center of the SNR G127.1+0.5 was initially viewed as a possible collapsed stellar remnant of the SN that produced the surrounding shell of radio emission.  Given the exceedingly low probability {\it{a priori}} of finding such an object at the very center of a SNR, this source was investigated and found to be a distant radio bright galaxy \citep{KC1978}}, 
the combination of finding a high latitude SNR so close to and in rough alignment with M31, whose only portion of its shell structure is dominated by [\ion{O}{3}] emission which faces away from the Galactic plane would seem to require an unusual combination of chance occurrences.

Finally, we considered whether SDSO might be an interstellar filament associated with an exceptionally large Galactic SNR or ISM bubble on the scale of several 10's of degrees in size. Filamentary structures in the ISM are not uncommon and can span a wide range of physical and density scales.  (See \citealt{Hacar2022} for an in-depth review of Galactic ISM filaments.)  Although interstellar filaments are commonly seen in radio, infrared and H~I studies, optical/UV detections of seemingly isolated filaments are rare. 

One notable exception is the straight and narrow H$\alpha$ filament some $2.5\degr$ long and located at Galactic latitude $+38\degr$ \citep{McCullough2001}. This was subsequently discovered to be part of a much larger $30\degr$ long filament best seen in GALEX images \citep{Bracco2020}. Other examples are two $\sim 2\degr$ long synchrotron filaments also detected in H$\alpha$ \citep{West2022}.
Explanations of such filamentary structures have included relic fragments of a very large and nearby SNR, a low density stellar jet, a superbubble or even parts of the Local ISM Bubble. However, none of these interstellar filaments exhibit bright [\ion{O}{3}] but weak H$\alpha$ emission.  None are as broad and diffuse in appearance as SDSO.

\subsubsection{A Nearby Planetary Nebula}

A sharply defined and curved optical nebula lacking strong radio emission and exhibiting some filament-like structure raises the possibility of SDSO being a Galactic PN, albeit an unusually large and close one. SDSO's morphology shown in Figures 1 and 2 bears some resemblance to that seen in a few PNe that exhibit optical emission largely on one side. The reason for their non-circular shape is their interaction with the ISM  
\citep{Tweedy1996,Ali2000,Sabin2010,Sabin2012,Weidmann2016,Frew2016}. 

Nonetheless, a PN nature for SDSO also seems doubtful. 
While it well established that PN can exhibit strong [\ion{O}{3}] emission with values of I(4959+5007)/I(H$\alpha$) in excess of 5 (e.g., the Ring Nebula; \citealt{Hawley1977}),
[\ion{O}{3}] emission is normally more concentrated to the center of the PN and close to the CSPN than that of H$\alpha$ and [\ion{N}{2}]. 
Yet we find no correlated H$\alpha$ + [\ion{N}{2}] emission similar in strength to that of SDSO's [\ion{O}{3}] in our images. 

The morphology, curvature, and size of the arc would require the projected radius of the nebula to be at least $2\degr$. This makes the linear radius 
$r_{\rm PN} > 0.035\times D$ with $D$ being the distance to the nebula. 
SDSO would easily be the largest PN in terms of angular size, dwarfing Sh~2-216's record-setting radius of $0.86\degr$. Sh~2-216, which is also currently one of the closest known PN ($D_{\rm GAIA} = 128$ pc), 
with an estimated age of 660,000 years \citep{Ziegler2009}, is among the oldest and the third largest PN with a radius of  1.92 pc and a central star with M$_{\rm G} = 7\fm08$ \citep{Smith2015}, 
just behind Ton~320's radius of 2.69~pc (central star M$_{\rm G} = 6\fm95$) and the gigantic nebula Alves 1 (PN G079.8-10.2) with a radius of 5.74~pc (M$_{\rm G} = 7\fm19$) (all using the new GAIA distances).
 With a proposed radius of at least 2 to 3 degrees, this thus would position SDSO within 100\,pc at most, but most likely below 50\,pc.

In the largest homogeneous catalog of planetary nebulae giving statistically derived diameters by \citet{Frew2016} there are 74 PNe larger than 1\,pc. From them, we selected those 38 having blue central stars detected in GAIA. As shown in \citet{KimeswengerGAIA2018}, only central stars with ($m_{BP} - m_{RP}) < 0$ have reliable GAIA distances. In the quick-look spectra, where available, of the Hong Kong/AAO/Strasbourg H$\alpha$ \citep[HASH,][]{HASH} planetary nebula database,
none of them has a line ratio of [OIII](4959+5007)/H$\alpha$ $>$ 1.
This sample of the oldest central stars with the largest known PNe has a mean absolute magnitude in the GAIA photometric system M$_{\rm G} = 7\fm08 \pm 1\fm03$ with the faintest to be M$_{\rm G} = 9\fm58$. 

This means that a potential central star for SDSO lying at the largest potential distances would have a typical apparent magnitude  m$_{\rm G} \approx 9\fm6$,
with an extreme limit of m$_{\rm G} \leq 12\fm0$.
An examination of stars within a radius of 10\degr\ around M31 resulted in 598 stars with distances below 110\,pc (note: to avoid Kerker bias the volume was stretched slightly), and magnitudes below 12\fm0 listed in GAIA DR3. Only two of them have a ($m_{BP} - m_{RP}) < 0$. None of them is a white dwarf. The most recent catalog of nearby white dwarfs by \citet{WD_2023} lists 88 WD stars within this field. By far the hottest star is FBS 0050+358 with only 26kK. This finding corresponds well to the spectroscopic surveys collected recently with LAMOST \citep{LAMOST2021,LAMOST2023} and the data collection by \citet{Geier2020}.

For expansion velocities of $30 - 50$ km s$^{-1}$, typical for old planetary nebulae, the expansion age for a PN is 
$\tau_{\rm exp} > (10^3~{\rm yr}) D_{\rm pc}$.
A complete grid of PNe models using CLOUDY \citep{Ferland2013} along the evolutionary tracks for stars with initial masses from 0.9 to 3 M$_\odot$ \citep{Miller-Bertolami2016} was calculated. This showed that only central stars with effective temperatures above 110kK and luminosities above 800 L$_\odot$ can give an excess of [OIII](4959+5007)/H$\alpha$ $>$ 2 at radii larger than 10$^{18}$ cm ($= 0.32\,$pc). However, even at a close distance of 5 pc, the dynamic age eliminates all progenitors with masses above 1.5\,M$_\odot$ using the ages along the evolutionary tracks. Moreover, none of the central stars of the sample of large old PNe derived above reaches this luminosity-temperature domain in the HRD. 
We conclude that such a luminous white dwarf, even if masked by a main sequence star in a binary system \citep[e.g., UCAC2 46706450;][]{Werner2020}, would not be missed at this low distance and the complete coverage of the region by GALEX.

Similar arguments eliminate the possibility of other ionizing blue stars. Blue main sequence stars or luminous blue variables (LBV) all have at least an order of magnitude higher luminosities.  Moreover, as shown in radius surface brightness diagrams, compact \ion{H}{2} regions and massive star ejecta (MSE) are at least an order of magnitude brighter in H$\alpha$ than all PNe of similar sizes 
\citet[e.g., their Fig.\ A1 in ][]{Frew2016}. However, we would still be missing the detection in the hydrogen lines with the same morphology as our [O III] arc.

\begin{deluxetable*}{lcccc}[ht]
\tiny
\centerwidetable
\tablecolumns{5}
\tablecaption{Comparisons of Likelihoods of Various Origin Scenarios for the SDSO Nebula }
\tablewidth{0pt}
\tablehead{\colhead{SDSO's}   &  \multicolumn{3}{c}{\underline{~~~~~~~~~~~~~~~~~~~~~~~~~~Galactic Nebula ~~~~~~~~~~~~~~~~~~~~~~}}        &  \colhead{\underline{~~~~~~~M31 Nebula~~~~~}}   \\
           \colhead{Properties}  &  \colhead{SN Remnant}      & \colhead{Planetary Nebula}   & \colhead{Stellar Bow Shock}   & \colhead{Halo-CGM Shock}    }
\startdata
  V$_{\rm Hel}$ $\sim 0$ km s$^{-1}$  & likely    & likely    & likely    & unlikely \\
  Majority of nebula with $[$\ion{O}{3}$]$/H$\alpha$ $>$ 5   & possible   & unlikely  & unlikely  & possible \\
  Absence of any associated H$\alpha$ emission   & unlikely  & unlikely  & unlikely & possible \\
  Percentage of known nebulae with Dia. $\geq 1.5\degr$ & $<$10\%$^{a}$ & $<0.003$\%$^{b}$  & $0$\%$^{c}$ & \nodata \\ 
  Probability of nebula at $|b| > 20\degr$     &  very low     & very low     &  very low    & 100\%     \\
  Probability of nebula within $ 2\degr$ of M31 &  very low     & very low     & very low     & 100\%     \\
  Probability of an one-sided nebula   & low   & low  & high   & 100\%    \\
\enddata
\tablenotetext{a}{\citet{Green2022} SNR catalog.}
\tablenotetext{b}{\citet{Frew2016} on-line PN catalog.}
\tablenotetext{c}{\citet{VanBuren1988}, \citet{Cox2012}, \citet{Decin2012}, \citet{Peri2015}. }
\label{Table_1}
\end{deluxetable*}

\subsubsection{A Bow Shock Nebula}

The curvature of the SDSO nebula suggests that it might be an interstellar 
bow shock preceding a fast-moving star in a nearby Galactic source. Strong shocks will be produced by stars moving at velocities (40--70~\kms) through the
ISM \citep{Shull2023} in excess of the sound speed $c_s$ (or magnetosonic wave speed $v_m$).  Fast-moving OB stars with strong winds have observed arc-like features of 
parsec size \citep{VanBuren1988,Mackey2016}. An O-star 
moving at velocity $V_* = (40~{\rm km~s}^{-1})V_{40}$ would produce a wind 
termination shock of the approximate size,  
 \begin{eqnarray}  
r_b &\approx& \left[\frac {\dot{M}_w V_w}{4 \pi \mu n_{\rm H}} \right]^{1/2} 
   V_*^{-1} \nonumber \\
  &\approx&  (1.0~{\rm pc}) \, \dot{M}_{-6}^{1/2} n_{\rm H}^{-1/2} 
   \left( \frac {V_w}{1000~{\rm km~s}^{-1}} \right)^{1/2} V_{40}^{-1}   \; 
\end{eqnarray} 
Here, we express the stellar mass-loss rate as 
$\dot{M}_w = (10^{-6}~M_{\odot}~{\rm yr}^{-1}) \dot{M}_{-6}$ and equate the wind ram pressure, $\dot{M}_w V_w / 4 \pi r^2$, with the pressure, $\rho_{\rm ISM} V_*^2$, of the ISM of mass density $\rho_{\rm ISM} = 1.4 m_{\rm H} n_{\rm H}$.  
At a distance $d \approx 100$~kpc, the angular size would be 
$\theta_b = r_b/d \approx 10^{-2}$ rad ($0.57\degr$), considerably smaller than 
the implied radius of the SDSO arc.

A bow shock could explain SDSO's limited extent and its one-sided nebula appearance.  SDSO's bright [\ion{O}{3}] emission relative to that of H$\alpha$ could signal a high stellar wind speed and/or source space velocity. 
Its broad morphology might arise from the projection of overlapping shock fronts 
due to variations in a star's mass-loss history.
However, we view it unlikely that SDSO represents a stellar bow shock nebula. 
The huge angular size of SDSO would require a nearby star with an enormous 
mass-loss rate that has somehow escaped prior identification.

\begin{figure*}[ht]
\begin{centering}
\includegraphics[angle=0,width=18.0cm]{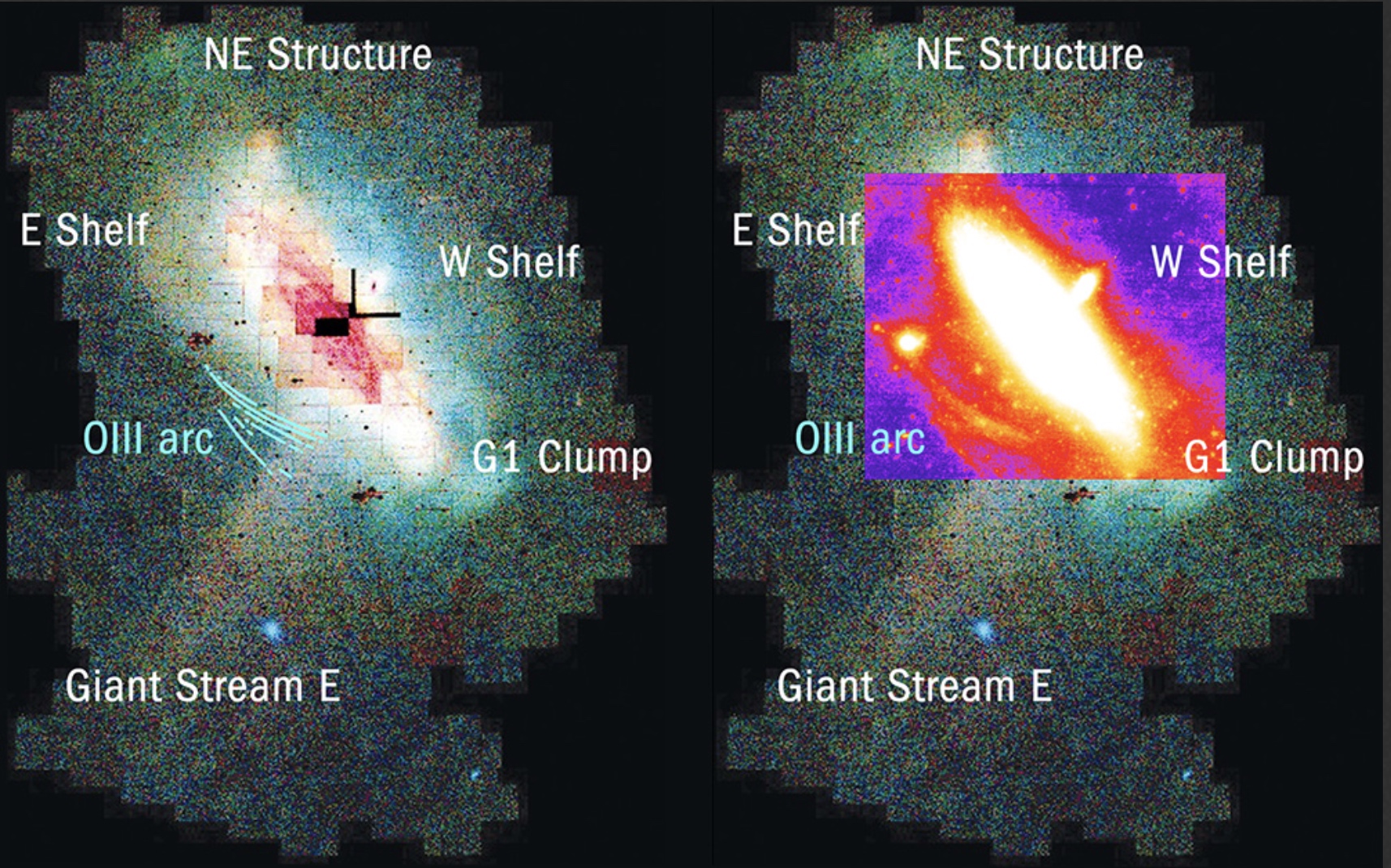} 
\caption{Location of SDSO relative to M31's stellar streams.
Left Panel: Figure adapted  from \citet{McConnachieINGN} 
with SDSO marked by several blue streaks. 
Right Panel:  Color insert of our [\ion{O}{3}] image 
overlaid on the same figure.  SDSO's projected position lies in a region 
in between M31's stellar streams with its westernmost extent matching 
at the edge of the Giant Stream East.
\label{fig12}
}
\end{centering}
\end{figure*}

Optical and infrared emission bow shocks are seen in several different 
astrophysical settings, including mass loss from high mass OB stars of which $\simeq25$\% are runaway stars \citep{VanBuren1988,Blaauw1993,Boden2018,Chick2020},
red supergiants \citep{Noriega1997,Meyer2014},
nova-like cataclysmic variables \citep{Castro2021},
pulsars \citep{Kulkarni1988,Brown2014}, 
and AGB carbon stars \citep{Libert2007, Weidmann2023}.
Unlike the highly curved bow shocks seen around pulsars, stars with space velocities
of a few 10's of km s$^{-1}$ lead to much less curved nebulae like that of SDSO, not unlike that seen in the infrared nebula around the runaway O9.5~V star Zeta Oph \citep{Green2022}.

In all the cases above, the source of the wind lies close to the bow shock nebula and at the apex of the bow shock. However, ignoring the lack of expected curved bow shock geometry in SDSO, we could not identify an obvious stellar source that could generate the observed nebula.
The nearest bright optical and GALEX FUV source to SDSO is the A0~E star HD~4336 (V = 9.01). With a Gaia DR3 parallax of 3.02 mas implying a distance of $\simeq$330 pc, its luminosity, and stellar winds are far too weak to produce a photoionized bow shock with the dimensions for SDSO of around 10 pc if at a distance of around 300 pc.
A bright FUV source projected farther away from SDSO is the symbiotic binary EG Andromedae (V = 7.22, M2.4 III + WD) 
(see Fig.\ 8) which has an estimated mass-loss rate of 
$\dot{M}_w \sim 10^{-6}$ M$_{\odot}~{\rm yr}^{-1}$. However, if SDSO was a bow shock associated with EG~And, the star's Gaia DR3 estimated distance of some $600 \pm 12$ pc would imply SDSO's physical size of roughly 15 pc, an order of magnitude larger in scale to the largest stellar bow shocks which have typical dimensions around 1 pc.

\begin{figure*}[ht]
\begin{centering}
\includegraphics[angle=0,width=8.0cm]{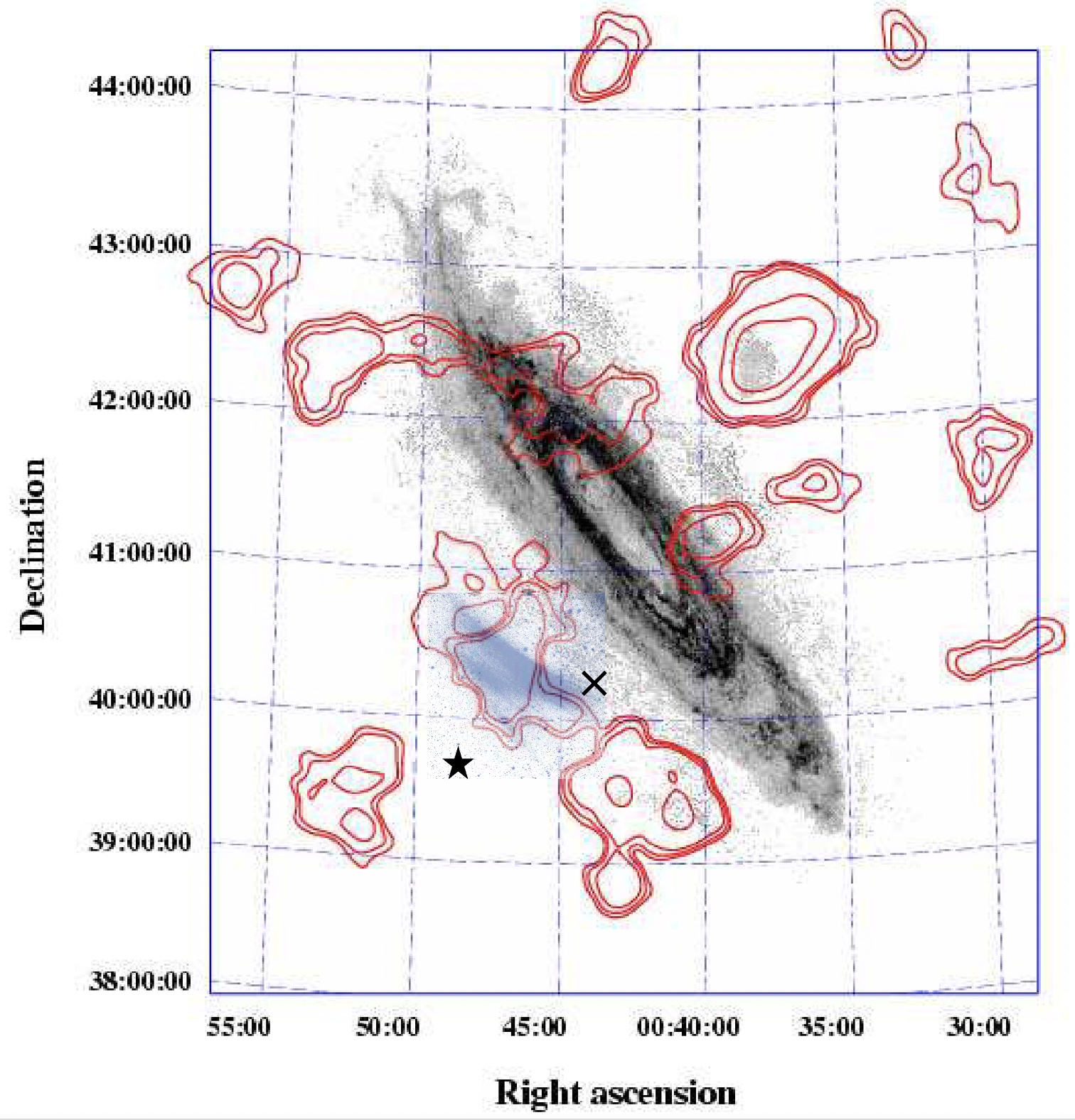} 
\includegraphics[angle=0,width=9.15cm]{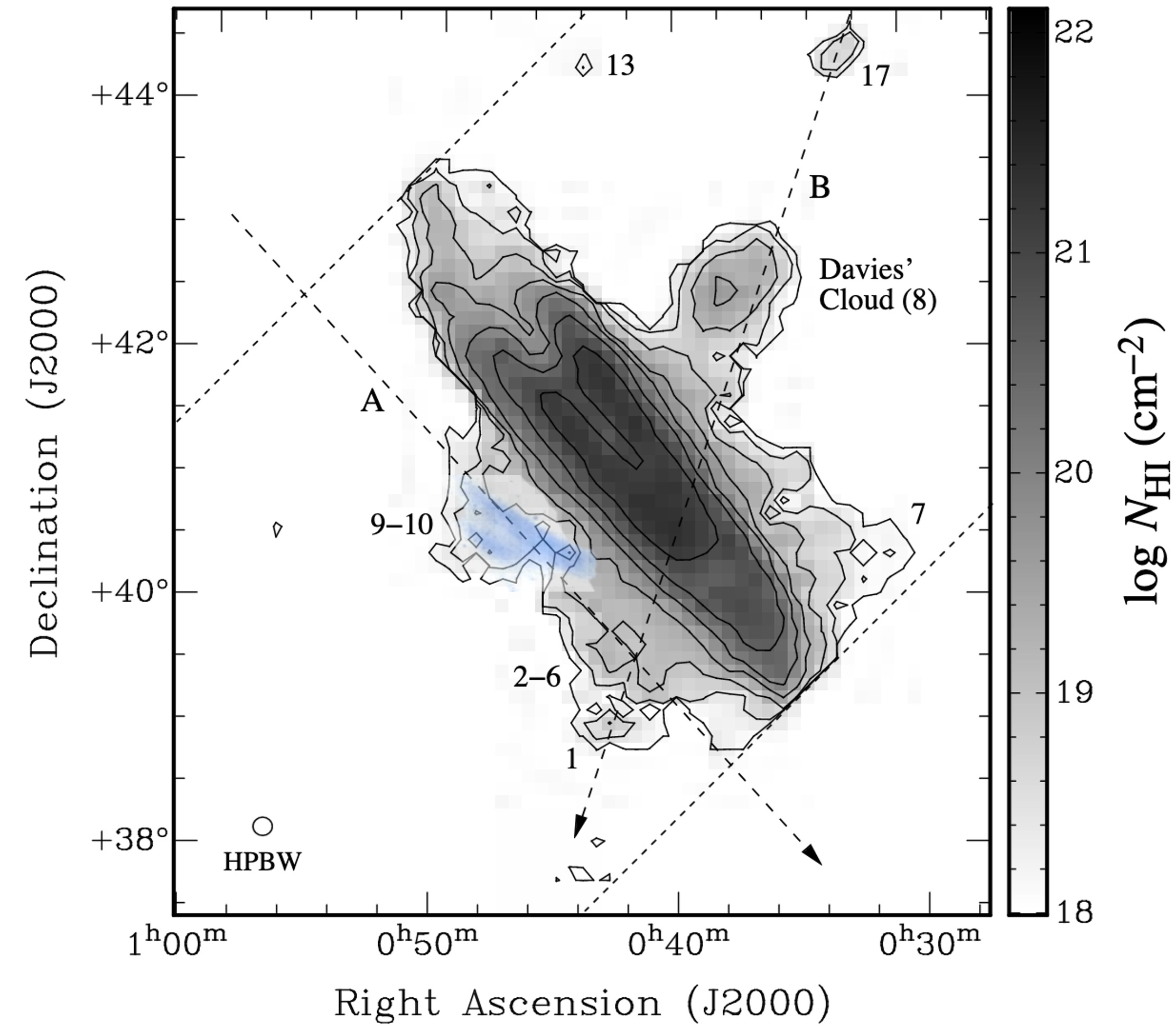} 
\caption{Location of SDSO projected onto maps of high-velocity H~I clouds near M31.
Left panel: Map from \citet{Thilker2004,Thilker2005a} with overlay of SDSO's [\ion{O}{3}] emission shown in blue. 
The positions of quasars near SDSO studied by
\citet{Rao2013} and by \citet{Lehner2015} are indicated
by $\times$ and $\bigstar$, respectively. SDSO's projected position coincides with an H~I cloud exhibiting an elongated morphology similar to SDSO's NE-SW appearance. 
Right panel: H I survey map of HVCs around M31 from \citet{Westmeier2008} with
SDSO's projected location again shown in blue.
 \label{fig13}
}
\end{centering}
\end{figure*}

\subsection{Extragalactic: A M31 Nebula}

The discussions above led us to the viewpoint that a Galactic origin for SDSO is unlikely. This conclusion is supported by Table 1, where we list the likelihood of the three Galactic origin scenarios for explaining some of SDSO's observed properties. The combination of low or unlikely probabilities for each of the galactic scenarios considered, namely an unrecognized SN remnant, a planetary nebula, or a stellar bow shock, led us to consider instead a scenario where SDSO is physically related to M31 as suggested by the close projection of SDSO so near M31 in the sky despite the  difference in their heliocentric velocities.

Large emission nebulae around massive galaxies showing [\ion{O}{3}] emission lines have been detected before. The most famous case is that of Hanny's Voorwerp near the spiral galaxy IC~2497 \citep{Lintott2009,Keel2012}. However, such cases are related to the galaxy's AGN emission, something not the case for M31.
Despite displaying a radial velocity more in line with a Galactic origin, 
the [{\ion{O}{3}] arc's projected location so close to M31 with a lateral structure nearly parallel to M31's disk naturally suggests a connection with M31. If the arc were physically near M31, it would have an extension of more than 100,000 light-years and would contain a substantial amount of matter.

The arc's average offset $\sim 1.2^{\circ}$ on the sky from M31's nucleus translates to a distance $(16.1~{\rm kpc}/\cos i)$ from the M31 nucleus, where $i \sim 72^{\circ}$ is the inclination angle of M31's disk plane. Thus, the filaments could be $\sim70$~kpc away from the M31 nucleus if they lie in the disk plane.  

An interaction of M31's halo with the circumgalactic medium would seem possible, in view of the numerous tidal debris streams that lace the M31 environs and its collection of small satellite galaxies: \citep{Ibata2001,McConn2003,McConn2005,McConn2018,vanderMarel2012,
Gilbert2007,Ferguson-Mackey2016,Gilbert2019,Fardal2007,Fardal2012}.
In considering possible M31-related scenarios, we examined stellar 
observations \citep{Dey2023} and N-body models \citep{Fardal2007} of stellar streams, particularly the GSS.  However, the modeled trajectory of the merging galaxy and the observed blueshifted velocities of the stars relative to M31 are inconsistent with the velocity measured for the filaments.  In the model, the merging galaxy approached M31 from the far side and passed back through M31.  The blue-shifted stars in the GSS are falling back toward M31. Thus, we were unable to find a consistent kinematic scenario connecting the GSS merger and the filaments.

Figure 12 shows the location of SDSO relative to the major M31 streams close to the M31 disk. SDSO can be seen as located east of the so-called Giant Stellar Stream (GSS) \citep{Ibata2001,Dey2023}, with its southwestern extent roughly coincident with GSS's sharp eastern edge. Because these stellar stream studies do not map the associated gas, it is possible that the gas seen in SDSO might mark the locations of tidal effects differently than  stripped stars, since gas is subject to dissipation and pressure gradients. Consequently, SDSO may be the signature of the M31 halo interacting with relic circumgalactic gas associated with ancient tidal disruption events.

In such a scenario, the [\ion{O}{3}] emission feature (SDSO) may be a thin shield of material in the M31 halo marking interaction with local HVCs or Local Group gas. The physical scale of such an interaction would be roughly the same as M31's disk, namely a few degrees, consistent with SDSO's angular size. It would also likely display a small concave curvature, again like  that of SDSO. M31's unusually high halo stellar mass fraction \citep{Trujillo2016} lends additional support to this general picture. 

In addition, the transverse velocity and proper motion vector of M31 vector points southeast of M31's nucleus \citep{vanderMarel2008,Sohn2012,vanderMarel2012,Salomon2021}.  This places it perpendicular to and passing through the SDSO filaments.
In such an interaction model, the southeastern boundary of the M31's halo can be considered as the ``leading edge'' of M31's Local Group motion.

The coincidence of SDSO's projected location with M31's motion is consistent 
with the possibility of an interaction between M31 and its HVCs or circumgalactic medium (CGM). However, while a velocity of M31 relative to the Milky Way of around 100~km~s$^{-1}$ could generate a shock bright in [\ion{O}{3}], it is not obvious how such an interaction is consistent with the low ($-10\pm7$ km s$^{-1}$) heliocentric velocity that we measure.

Interestingly, some of the intergalactic gas around M31 is believed to be from a tidal encounter between M31 and M33 some 1-3 Gyr ago \citep{Blitz1999,Braun2004,Bekki2008}. This left a discontinuous stream of H~I clouds between the two galaxies as reported by \citet{Braun2004} and \citet{Thilker2004} and subsequently confirmed and investigated by \citet{Lockman2012},
  \citet{Westmeier2008}, \citet{Wolfe2013}, and \citet{Kerp2016}.

Figure 13 shows the location of high velocity clouds around M31 
reported by \citet{Thilker2004,Thilker2005a} and \citet{Westmeier2008}
thought to be the equivalent to the Milky Way's HVCs.
The location of SDSO  matches one of these HVCs southeast of the M31 nucleus with a projected location coincident with one of these clouds and exhibiting a morphology similar to that cloud's NE-SW elongation. However, the heliocentric velocity of this HVC is $-175$ km s$^{-1}$, far from SDSO's $-10$ km s$^{-1}$ value. 

Such H~I studies of HVCs near M31 typically only survey radial velocities from $-100$ to $-600$ km s$^{-1}$ and are hence blind to low velocity gas. However, a study of H~I clouds near M31 by \citet{Kerp2016} covered velocities down to $-25$ km s$^{-1}$ with a angular resolution of $10.8'$. While detecting no significant low velocity H~I gas at SDSO's location, the curved M31-M33 stream of HVCs aligns with SDSO's location off M31.

We note that \citet{Rao2013} and \cite{Lehner2020} analyzed low-dispersion UV spectra taken with the Cosmic Origins Spectrograph on {\sl Hubble Space Telescope} to explore 
the CGM around M31, using metal absorption lines toward background quasars. However, the velocities seen, relative to the core of M31 are fairly small. The locations of two of these quasars are shown in the left panel of Figure 13. The sight-line to one of the \citet{Rao2013} targets, 0043+4016, is coincident with very faint SDSO's [\ion{O}{3}] emission along its northern limb (shown in the figure as an $\times$ symbol). However, strong line blending and low spectral resolution of the \ion{Si}{2} $\lambda$1260 and \ion{C}{4} $\lambda$1548 features limited a clear HVC detection.

A second target from the \citet{Lehner2015} list, RX\ J0048.3+3941, is located $1.9\degr$ from the M31 core and about $20'$ outside our arc-like features (shown in the figure as a $\bigstar$ symbol). \citet{Lehner2020} detected absorption lines from ionized metals (\ion{C}{2}, \ion{Si}{3}, \ion{Si}{4}, \ion{C}{4}) at LSR velocities between $-170$ km~s$^{-1}$ and $-300$ km~s$^{-1}$, similar to that of the H~I cloud.  They interpret these ionized absorption features as gas in the CGM of M31.  However, as shown in Figure 13, the quasar sight line does not pass through the 21-cm contours of the H~I cloud.  On the other hand, Galactic HVCs and clouds in the CGM are expected to be ionized and could extend well beyond the H~I contours. 

While positional coincidences of SDSO with M31's circumgalactic H~I clouds are suggestive, we do not know the 3D location of SDSO relative to the high (or low) velocity H~I emission clouds.  Assuming that SDSO represents shocked halo gas or 
CGM at roughly M31's 770~kpc distance, we can estimate some of its basic properties.  For filaments extending over 900 arcmin$^2$ ($15' \times 60')$ with surface 
brightness $I_{\rm arc}$ scaled to 0.1 Rayleigh, the radiated luminosity is 
$(1.7 \times 10^{38}~{\rm erg~s}^{-1}) (I_{\rm arc} / 0.1~{\rm R})$.  
We assume that the [\ion{O}{3}] is the dominant ionization state of oxygen in collisionally excited gas at $T \approx 10^4$~K, with $\lambda 5007$ 
collision strength $\Omega_{12} = 2.19$, hydrogen density 
$n_H \approx 10^{-2}$~cm$^{-3}$, and metallicity $Z$ relative to the
solar oxygen abundance 
$(n_{\rm O}/n_{\rm H})_{\odot} = 5.62\times10^{-4}$. From the observed 
surface brightness, we compute a volume emission measure,
\begin{equation}
   {\rm EM} \equiv n_H^2 V_{\rm arc} = (6.55\times10^{61}~{\rm cm}^{-3})
      \left[ \frac {I_{\rm arc}} {0.1~{\rm R}} \right]
      \left[ \frac {Z}{Z_{\odot}} \right]^{-1} \; , 
\end{equation}
and a mass ($1.4 n_{\rm H} m_{\rm H} V_{\rm arc}$) of 
\begin{equation}
  M_{\rm arc}  \approx (8 \times 10^{6}~M_{\odot}) 
      \left[ \frac {I_{\rm arc}} {0.1~{\rm R}} \right]
       \left[ \frac {Z}{Z_{\odot}} \right]^{-1} 
      \left[ \frac {0.01~{\rm cm}^{-3}} {n_H} \right] \; .
\end{equation}
We scaled the uncertain density in the filaments to 
$n_{\rm H} = 0.01~{\rm cm}^{-3}$, typical of Galactic HVC densities 
\citep{Collins2007}. This density is consistent with a rough estimate 
(0.004~cm$^{-3}$) from the emission measure, assuming that the filaments 
have a depth of $15'$ comparable to their radial extent on the sky. 
Such densities would allow the gas to cool on dynamic time scales for
100 km s$^{-1}$ tidally stripped outflows across 50--100 kpc distances.  

\section{Conclusions}

The presence of an exceedingly faint nebulosity (SDSO) strong in [\ion{O}{3}] emission 
within 1.2$\degr$ of M31 was an unexpected discovery made through the addition of hundreds of exposures taken by amateur astronomers 
using small telescopes equipped with narrow, high throughput passband filters and sensitive digital detectors. 
Here we presented images that better defined the nebula's extent and structure, plus radial velocity results from low-dispersion spectra. A summary of our main results and findings follows:

1) Deep [\ion{O}{3}] images show SDSO to be composed of diffuse emission streaks up to $2\degr$ in length. Deep H$\alpha$ images reveal no strong coincident emission, suggesting a high ratio of I([\ion{O}{3}] $\lambda\lambda$4959,5007)/I(H$\alpha$).

2) We find no other [\ion{O}{3}] emission nebulosity as bright as SDSO within several degrees of M31, and no filamentary H$\alpha$ emission that might be connected to SDSO. 
We also find no far UV emission coincident with SDSO in GALEX images.

3) Long slit, low-dispersion optical spectra taken at two locations along the arc's northern edge reveal faint [\ion{O}{3}] 
emission matching the location and extent of emission seen in our [\ion{O}{3}] images. Because this emission vanishes outside the nebulosity seen in our images, we are confident that we have detected emission from SDSO. We estimate a  heliocentric velocity of  $-9.8 \pm 6.8$ km s$^{-1}$.

4) The surface brightness of SDSO's brighter regions ranges from
$(2.5 - 4.0) \times 10^{-18}$ erg s$^{-1}$ cm$^{-2}$  arcsec$^{-2}$.
These values are consistent with the earlier estimate reported in \citet{Drechsler2023}. SDSO's brighter regions have a surface brightness of $\sim 0.55$ Rayleigh, with the faintest regions in our [\ion{O}{3}] images detected at $\sim 0.2$ Rayleigh.

5) Because of SDSO's low radial velocity, we have considered various possible Galactic origins for this strong [\ion{O}{3}] but weak H$\alpha$ emission nebula.  We conclude that a supernova remnant, nearby planetary nebula, or stellar bow shock nebula explanations for its origin are all unlikely. Instead, we favor an extragalactic origin involving interaction of M31's outer halo with a circumgalactic high velocity cloud leading to shock emission. This cloud may be related to a large and ancient stream of discrete H~I clouds between M33 and M31 that might be the result of a past tidal disruption event.

\medskip

We note that our path to favoring an extragalactic halo--cloud interaction scenario was not a straight one. Initially, SDSO's low radial velocity seemed to be clear evidence for it being a Galactic nebula. But no Galactic nebula scenario seemed to fit, whereas SDSO's emission properties and positional coincidences in line with an extragalactic scenario were hard to dismiss. These included its visibility in [\ion{O}{3}] but not in H$\alpha$, its broad and largely diffuse emission structure despite its strong [\ion{O}{3}] line emission, its apparent isolation from other H$\alpha$ and [\ion{O}{3}] nebulae in this region of the sky, its alignment with and proximity to M31's disk, the lack of any potential ionizing Galactic sources in this direction,
its location southeast of M31's nucleus consistent with the galaxy's transverse velocity and proper motion vector with its NE-SW structure nearly perpendicular to this motion, and its projected coincidence with an H~I cloud near M31 with which it shares a similar NE-SW alignment. None of these by themselves are very persuasive, but viewed together they make a good case for an extragalactic origin.

There are a number of follow-up observations that could help resolve the true nature and origin of the SDSO nebulosity.
Although we are confident in our radial velocity measurement,  our result and that of \citet{Amram2023} were obtained with low dispersion spectra. Higher resolution spectra would be especially useful to investigate its radial velocity across SDSO's whole structure.  Spectra covering several emission lines including [\ion{O}{1}] $\lambda\lambda$6300,6364, 
 [\ion{O}{2}] $\lambda\lambda$3726,3729, [\ion{O}{3}] $\lambda\lambda$4959,5007, 
 [\ion{N}{2}] $\lambda\lambda$6548,6583,
 [\ion{S}{2}] $\lambda\lambda$6716,6731, H$\alpha$, and
 H$\beta$ could provide valuable data on the nebula's density and ionization state. Sensitive radio observations could also provide a test for the shock scenario as the origin of the SDSO's strong [\ion{O}{3}] emission through the detection of nonthermal radio emission.  Finally, sensitive and high resolution 21 cm data of the H~I emission cloud coincident with SDSO and covering the velocity range of $-10$ to $-100$ km s$^{-1}$ could investigate the presence of low velocity gas at SDSO's exact location as we have proposed.  

\bigskip

We thank Arjun Dey, Mark Fardal, Gary Ferland, Andrew Fox, and Wolfgang Reich for helpful comments and conversations about the SDSO nebula and M31's stellar streams. We are grateful to Eric Galayda and the entire MDM staff for making the optical spectral observations possible. We also thank Andrew Fryhover, Vicent Peris, and Alicia Lozano for generously making available their deep H$\alpha$ images of M31.
This work made use of the Simbad database, NASA's Skyview online data archives, the Max Planck Institute for Radio Astronomy Survey Sampler, and the Gaia EDR3 catalog.

\facilities{MDM Observatory (OSMOS), and various privately owned and operated telescopes }

\software{PYRAF \citep{Pyraf2012}, Astropy v4.0 \citep{Astropy2013}, ds9 \citep{DS9}, L.A.\ Cosmic \citep{vanDokkum2001}, ESO-MIDAS \citep{MIDAS} }

\bibliography{biblio}

\end{document}